\documentclass[superscriptaddress,prx,twocolumn,nofootinbib,citeautoscript,longbibliography,notitlepage]{revtex4-1}
\usepackage{graphicx}
\usepackage{dcolumn}
\usepackage{bm}
\usepackage{color}
\usepackage{amsmath}
\usepackage{tabularx,graphicx}
\usepackage{epstopdf}
\usepackage{latexsym}
\usepackage{amssymb}
\usepackage{amsmath}
\usepackage{color, colortbl}
\usepackage{psfrag}
\usepackage{bbm}
\usepackage{booktabs}
\usepackage{bm}
\usepackage{titlesec}
\usepackage{dsfont}
\usepackage{feynmp}
\usepackage{slashed}
\usepackage{multirow}
\usepackage[tight]{subfigure}
\usepackage{comment}
\usepackage{float}
\usepackage{diagbox} 
\usepackage{makecell}
\usepackage{array}
\usepackage[papersize={8.5in,11in}]{geometry}
\usepackage{color}
\usepackage{fmtcount} 
\definecolor{darkblue}{rgb}{0.,0.,0.4}
\definecolor{darkred}{rgb}{0.5,0.,0.}
\definecolor{BlueViolet}{RGB}{138,43,226}
\definecolor{SkyBlue}{RGB}{30,144,255}
\definecolor{DarkGreen}{RGB}{0,100,0}

\usepackage[pdftex,colorlinks=true,linkcolor=darkblue,citecolor=blue,urlcolor=darkred]{hyperref}

\renewcommand{\epsilon}{\varepsilon}

\allowdisplaybreaks[3]

\usepackage{cases}

\allowdisplaybreaks[4]

\UseRawInputEncoding

\begin{document}



\title{Symmetry-protected isolation of the quantum metric in chiral type-II semi-Dirac semimetals}

\author{Wen-Hao Bian}
\altaffiliation{whbian@smail.nju.edu.cn}
\affiliation{Department of Physics, Tianjin University, Tianjin 300072, P.R. China}
\affiliation{School of Physics, Nanjing University, Nanjing, Jiangsu 210093, P.R. China}

\author{Jing Wang}
\altaffiliation{jing$\textunderscore$wang@tju.edu.cn}
\affiliation{Department of Physics, Tianjin University, Tianjin 300072, P.R. China}
\affiliation{Tianjin Key Laboratory of Low Dimensional Materials Physics and Preparing Technology,
Tianjin University, Tianjin 300072, P.R. China}

\date{\today}


\begin{abstract}

Quantum geometry, comprising the Berry curvature (BC) and the quantum metric (QM),
underpins a wide array of topological and transport phenomena. However, in most solid-state
platforms, BC and QM signals are inherently entangled, making their experimental separation
highly challenging. Here, we demonstrate that the chiral limit of type-II semi-Dirac semimetals
provides a symmetry-protected mechanism to cleanly isolate the QM. At the chiral limit where the
spin-orbit gap $m_z=0$, chiral and $\mathcal{PT}$ symmetries strictly suppress the BC,
its dipole (BCD), and the shift current, leaving the QM and its dipole (QMD) as the sole sources
of the optical and nonlinear Hall responses. This pure-QM regime exhibits distinct experimental
fingerprints, including a divergent longitudinal optical conductivity and a QMD-induced nonlinear
Hall conductivity that exhibits a pronounced singularity as the chemical potential approaches the
nodal point. Turning on a finite $m_z$ reactivates the BC channels, transitioning the system
from a pure-QM phase into a mixed geometric phase. By deriving a scaling law for the nonlinear
Hall conductivity, we establish a direct experimental strategy to separate the mixed Drude, BCD,
and QMD contributions based on their distinct relaxation-time dependencies. Our findings identify
chiral type-II semi-Dirac semimetals as highly tunable platforms for isolating and probing intrinsic quantum
geometric effects.

\end{abstract}


\maketitle


\section{Introduction}

The geometric structure of Bloch wavefunctions has emerged as a unifying principle in modern condensed-matter physics~\cite{Liu2025NSR,Gao2026RMP}. This internal geometry is fully captured by the quantum geometric tensor (QGT),
which has become one of the most active research frontiers~\cite{Provost1980CMP,Berry1984PRSA,Liu2025NSR,
Gao2026RMP,Sodemann2015PRL,Sipe2000PRB,Das2023PRB,Gao2023SCI,Gao2023SCI,Wang2023Nature,Orenstein2021ARCMP}.
The QGT comprises the antisymmetric Berry curvature (BC) and the symmetric quantum metric
(QM)~\cite{Provost1980CMP,Berry1984PRSA}. While the BC encodes the phase information of the wavefunctions and
drives topological transport~\cite{Liu2025NSR,Gao2026RMP}, the QM measures the amplitude overlap
and defines the Fubini-Study distance between adjacent Bloch states in the Brillouin
zone (BZ)~\cite{Xiao2010RMP,Liu2025NSR,Gao2026RMP}.

Recently, the QM has garnered immense attention for its intimate connection to a variety of quantum-geometric responses,
including the QM dipole (QMD) nonlinear Hall effect, shift currents, singular interband optical absorption~\cite{Sodemann2015PRL,Sipe2000PRB,Das2023PRB,Gao2023SCI}, and the BC polarization-induced
intrinsic nonlinear Hall effect in inversion-symmetric systems~\cite{Das2024PRL}. A major breakthrough was the experimental
observation of the QMD-induced nonlinear Hall effect in topological antiferromagnetic $\mathrm{MnBi_2Te_4}$
heterostructures, confirming that the QM is a physically measurable quantity~\cite{Gao2023SCI,Wang2023Nature}.
However, a fundamental challenge remains in that Bloch eigenstates in nearly all known solid-state platforms
generically carry a nonzero BC alongside a finite QM within the same band subspace~\cite{Kang2025NP,Kim2025SCI,Sala2025SCI}.
Consequently, QM signals are inherently entangled with and often masked by concurrent BC contributions.

This entanglement is particularly problematic in intrinsic second-order nonlinear Hall conductivity,
which receives additive contributions from two independent geometric invariants, namely the Berry curvature
dipole (BCD) and the QMD~\cite{Yu2025NC,Park2026PRB}. Experimentally disentangling these two terms poses a
major obstacle because they transform identically under several common crystal symmetries and contribute to
the same transport observables~\cite{Zhu2025NC}. The current state-of-the-art isolation in $\mathrm{MnBi_2Te_4}$
relies on suppressing the BCD via time-reversal-breaking magnetic order~\cite{Gao2023SCI,Wang2023Nature}.
However, since this approach depends on extrinsic, sample-specific manipulation, it lacks broad applicability.
It is therefore highly desirable to find alternative material platforms where the BCD vanishes intrinsically
by symmetry, leaving the QMD as the exclusive geometric contribution.

Semi-Dirac semimetals (SDSMs) are a distinct class of low-dimensional band systems characterized
by highly anisotropic low-energy dispersions, which are linear along one momentum axis and quadratic
along the orthogonal axis, setting them apart from isotropic Dirac and Weyl fermions~\cite{Pardo2009PRL,
Banerjee2009PRL,Delplace2010PRB}. This unique anisotropy yields unconventional transport and topological
phenomena, such as direction-dependent mobility, anisotropic Landau quantization, and protected edge modes~\cite{Banerjee2012PRB,Xiong2023PRB,Sinha2020PRB}. Unlike type-I SDSMs, type-II SDSMs break emergent
Lorentz symmetry via tilted band cones, resulting in reshaped Fermi pockets and modified interband selection rules~\cite{Yan2023PRB,Elsayed2026PRR}. Previous theoretical studies have primarily addressed band topology,
Fermi-surface geometry, and Berry-phase transport in these systems~\cite{Banerjee2012PRB,Huang2015PRB,
Carbotte2019PRB,Oriekhov2022PRB}. However, the full QGT remains largely unexplored, particularly regarding
the separation of its BC and QM components. Crucially, the intrinsic anisotropic geometry of SD
Hamiltonians provides natural symmetry constraints that enable the decoupling of the symmetric (QM)
and antisymmetric (BC) parts of the QGT. This renders SDSMs as ideal candidates for isolating
individual quantum geometric responses.

Motivated by the above analysis, we investigate a type-II SDSM~\cite{Huang2015PRB,Elsayed2026PRR,Ghosh2026PRB}
and demonstrate that restoring chiral symmetry by tuning the spin-orbit gap $m_z$~\cite{Huang2015PRB,Elsayed2026PRR}
enables the clean isolation of the QM contribution. In principle, at $m_z=0$, the low-energy Hamiltonian
of type-II SDSM respects both chiral symmetry $\{\sigma_z,H\}=0$ and antiunitary $\mathcal{PT}=\sigma_x K$
symmetry~\cite{Bera2026arXiv}. This strictly suppresses the BC while preserves the singular QM. Consequently,
all BC-related responses vanish, leaving only pure QM contributions.

Specifically, for the optical responses, a finite $m_z$ induces distinct peaks in both the
linear optical conductivity and the nonlinear shift current at energies determined by the band gap.
In the chiral limit ($m_z=0$), the BC-driven optical Hall conductivity and the Berry-connection-driven
shift current exactly vanish, leaving the QM-driven longitudinal optical conductivity as the sole
surviving channel. This accordingly directly extracts the optical response signal from the pure QM channel.
In addition, we examine the nonlinear Hall response, which contains both Drude and quantum geometric contributions.
For a finite $m_z$, symmetry dictates that the BCD dominates the response in the $x$-direction,
while the Drude and QMD contributions appear in the $y$-direction. At $m_z=0$, the BCD contribution
vanishes, but the $y$-direction response remains a mixture of Drude and QMD terms.
By employing scaling laws~\cite{Galitski2009PRB}, we provide a feasible method to experimentally
separate these two contributions, enabling the effective extraction of the pure QMD response.
This therefore establishes a highly tunable
platform for probing quantum geometry.

The remainder of this paper is organized as follows. In Sec.~\ref{SecII}, we introduce the low-energy Hamiltonian, derive the QGT, and analyze its components (the BC and QM). Section~\ref{SecIII} examines the linear optical conductivity and the shift current, while Sec.~\ref{SecIV} investigates the nonlinear response and scaling laws. Finally, we provide a brief discussion and summarize our conclusions in Sec.~\ref{SecV}.

\section{Microscopic model and quantum geometry}\label{SecII}

In this section, we construct the low-energy Hamiltonian for the type-II SDSM~\cite{Huang2015PRB,Elsayed2026PRR,Ghosh2026PRB,Bian2026arXiv}
and examine its underlying symmetries. By deriving the QGT~\cite{Ezawa2024PRB,Wiedmann2025PRB,
Liu2025NSR,Li2025arXiv,Bera2026arXiv,Villegas2025PRB}, we explicitly separate its BC and QM
contributions, which establishes the basis for the symmetry-protected isolation of the QM analyzed below.

\subsection{Low-energy Hamiltonian}

\begin{figure}
\centering
\hspace{-1cm}
\subfigure[$\delta_0<0$]
{\includegraphics[width=1.6in]{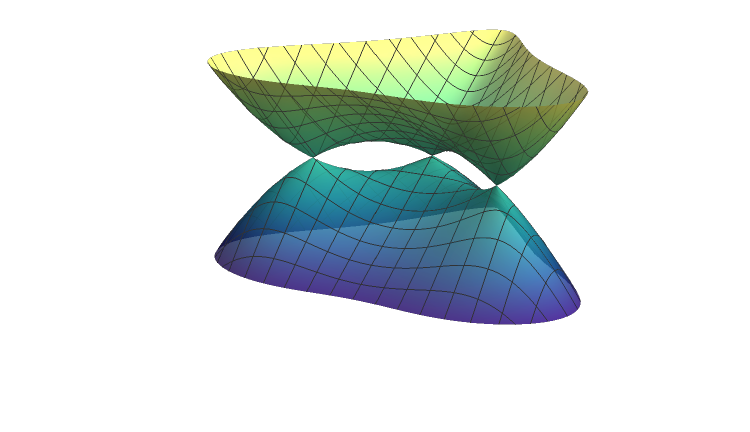}}
\hspace{-2cm}
\subfigure[$\delta_0=0$]
{\includegraphics[width=1.6in]{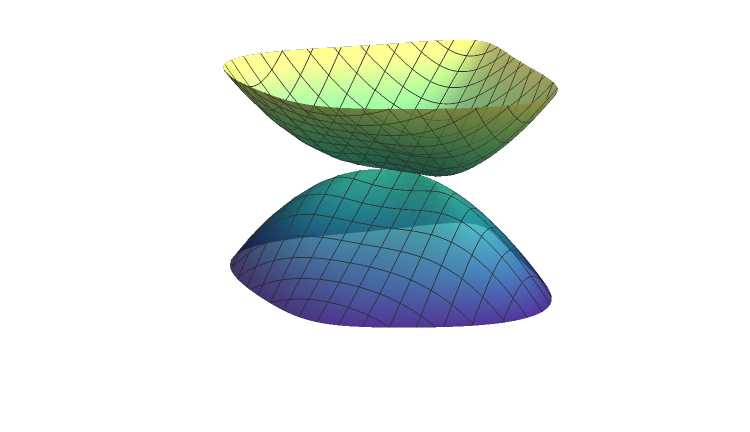}}
\hspace{-2cm}
\subfigure[$\delta_0>0$]
{\includegraphics[width=1.6in]{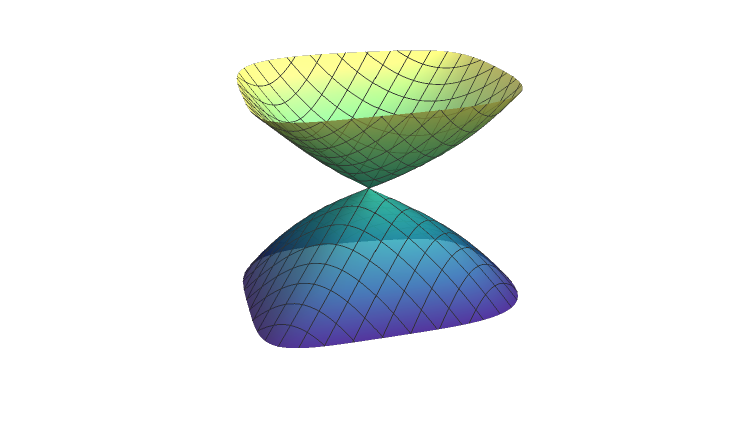}}
\vspace{-0.2cm}
\caption{(Color online) Low-energy dispersion $E(\mathbf{k})$ of the type-II SDSM with $m_z = 0$ in momentum space.
The panels show the band structure for (a) $\delta_0 < 0$, (b) $\delta_0 = 0$, and (c) $\delta_0 > 0$.}
\label{fig:E_vs_k}
\end{figure}

To capture the low-energy electronic structure of a two-dimensional type-II SDSM,
we employ the minimal effective Hamiltonian~\cite{Huang2015PRB,Elsayed2026PRR,Ghosh2026PRB,Bian2026arXiv}
\begin{eqnarray}
H(\mathbf{k})
&=&\mathbf{d}(\mathbf{k})\cdot\bm{\sigma},\label{Eq_H}
\end{eqnarray}
where $\mathbf{k}=(k_x,k_y)$ serves as the crystal momentum. For simplicity,
we introduce a pseudospin vector
\begin{eqnarray}
\mathbf{d}(\mathbf{k})\equiv
(\gamma_1 k_x^2-vk_y+\delta_0, \gamma_2k_xk_y, 0),\label{Eq_d_k}
\end{eqnarray}
and a Pauli-matrix vector $\bm{\sigma}\equiv(\sigma_x,\sigma_y,\sigma_z)$
acting in pseudospin space. Here, $\gamma_1$ denotes the inverse effective mass, $v$ the
velocity along the massless direction, $\gamma_2$ the cross-anisotropy term governing the type-II nature of the quasiparticles,
and $\delta_0$ the gap parameter.

The distinctive feature of type-II SDSMs is the off-diagonal $k_xk_y$ cross term in Eq.~(\ref{Eq_d_k}),
which mixes the two momentum components. This fundamentally distinguishes them from their type-I counterparts~\cite{Huang2015PRB,Elsayed2026PRR,Bian2026arXiv,Uchoa-Elsayed2026arXiv}, where the dispersion is quadratic
along one axis and linear along the perpendicular axis. In addition, this cross term modifies the Fermi
surface geometry, breaks specific intrinsic lattice symmetries, and gives rise to nonreciprocal band properties.

While the basic Hamiltonian~(\ref{Eq_H}) specifies the band dispersion, a complete description of the QGT
requires accounting for the Berry curvature. To this end, we introduce a spin-orbit coupling (SOC) mass term,
$\delta H = m_z\sigma_z$~\cite{Huang2015PRB,Elsayed2026PRR,Bian2026arXiv,Bian2026PRB}.
The full Hamiltonian is then described by the generalized vector $\mathbf{h}(\mathbf{k})$,
which extends the original vector $\mathbf{d}(\mathbf{k})$ by incorporating the SOC term,
\begin{equation}
\mathbf{h}(\mathbf{k})\equiv(h_1,h_2,h_3)=(\gamma_1 k_x^2-vk_y+\delta_0,\gamma_2k_xk_y,m_z).\label{Eq_H2}
\end{equation}
Here, it is worth highlighting that the non-zero $m_z$ component is essential for generating a finite
BC, as it breaks the symmetry constraints that would otherwise force the QGT to be purely real~\cite{Ahn-Nagaosa2020PRX,Kwon-Yang2024PRB}.
As will be shown below, this allows us to systematically trace the evolution of both the quantum metric
and the BC as a function of $m_z$.

\subsection{Band structure and symmetry analysis}\label{Sec_IIB}

Subsequently, we examine the band structure and underlying symmetries of the model.
By diagonalizing the Hamiltonian constructed from the vector in Eq.~(\ref{Eq_H2}),
we obtain the energy dispersion of the type-II SDSM as follows
\begin{eqnarray}
E^\pm(\mathbf{k})=\pm\sqrt{(\gamma_1 k_x^2-v k_y+\delta_0)^2+\gamma_2^2k_x^2k_y^2+m_z^2},
\end{eqnarray}
where $\pm$ corresponds to the conduction and valence bands, respectively, as shown in Fig.~\ref{fig:E_vs_k}.
The nodal structure is dictated by the sign of $\delta_0$. For $\delta_0>0$, a single node emerges at
$\mathbf{k}=(0,\delta_0/v)$. Conversely, for $\delta_0<0$, a Lifshitz transition yields three nodes
at $\mathbf{k}=(0,\delta_0/v)$ and $(\pm\sqrt{-\delta_0/\gamma_1},0)$~\cite{Volovik2016arXiv,Volovik2018PU}.

\begin{figure}
\centering
\vspace{0.5cm}
\includegraphics[width=\linewidth]{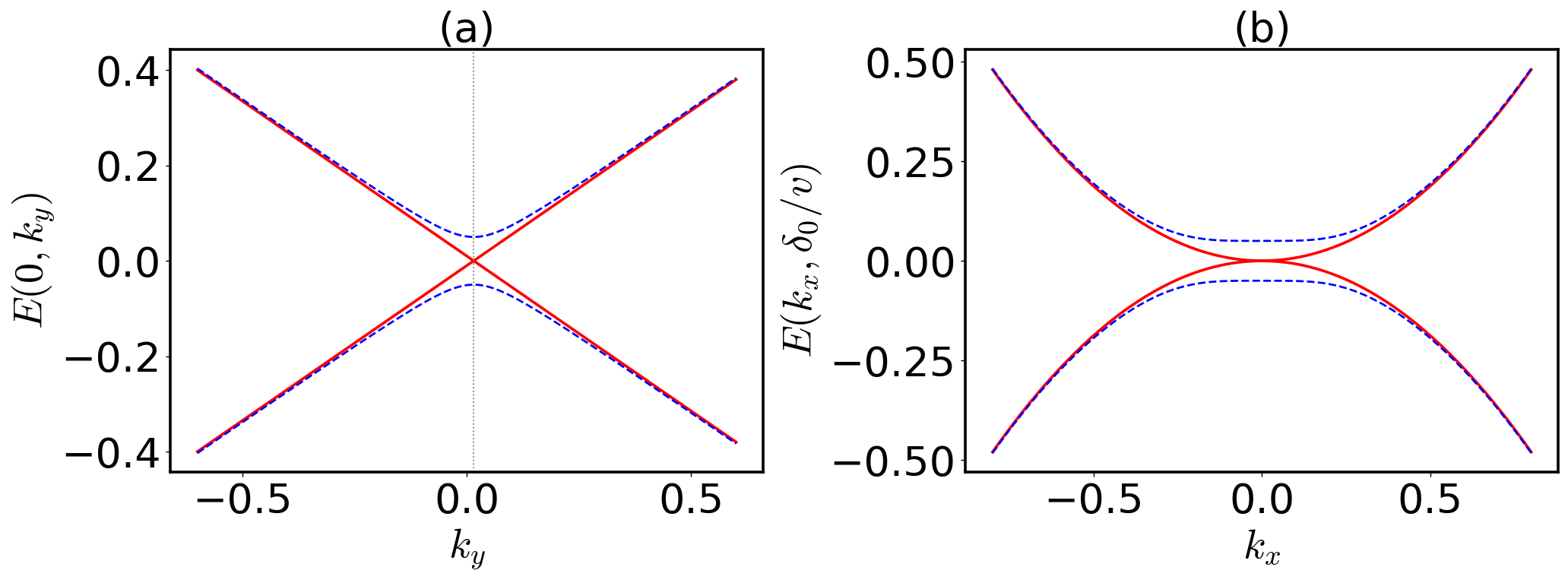}
\vspace{-0.5cm}
\caption{(Color online) Band dispersion around the nodal point for $\delta_0 > 0$
along (a) $k_y$ at $k_x = 0$ and (b) $k_x$ at $k_y = \delta_0/v$.
The red solid lines represent the gapless chiral limit ($m_z = 0$),
while the blue dashed lines show the gapped spectrum for a finite spin-orbit gap ($m_z = 0.05$).}
\label{fig:E_cut}
\end{figure}

Despite the differing number of nodes in these two cases, the underlying topological properties remain unchanged~\cite{Huang2015PRB,Elsayed2026PRR,Bian2026arXiv}. The minimum band gap is given by
$\Delta_{\rm min}=2\min_{\bf k}|{\bf h}(\mathbf{k})|=2|m_z|$, which closes at \(m_z=0\). Figure~\ref{fig:E_cut}
presents a representative illustration of how a finite $m_z$ modulates this gap for $\delta_0 > 0$.

Because quantum geometry is deeply constrained by symmetries of the Hamiltonian,
it is necessary to examine the symmetries of the type-II SDSMs.

We first consider the chiral limit where the spin-orbit gap vanishes ($m_z=0$).
In this limit, the type-II SDSM satisfies chiral symmetry,
\begin{eqnarray}
\{\sigma_z, H({\bf k})\} = 0,
\end{eqnarray}
which tightly binds the conduction and valence bands, resulting in particle-hole symmetry.
In addition, the system respects the antiunitary $\mathcal{PT}$ symmetry, defined as $\mathcal{PT}\equiv\sigma_x K$
where $K$ denotes complex conjugation~\cite{Bera2026arXiv}, such that
\begin{eqnarray}
(\sigma_x K)H(\mathbf{k})(\sigma_x K)^{-1}=H(\mathbf{k}).
\end{eqnarray}
This constraint forces the Berry curvature (BC) to vanish while leaving the QM
generally unrestricted~\cite{Li2025arXiv}.

When a finite spin-orbit gap $m_z$ is introduced, both the chiral symmetry and the $\mathcal{PT}$ symmetry are broken,
leading to the emergence of a nontrivial BC. Notably, for any $m_z$, the system intrinsically preserves an antiunitary mirror symmetry $\mathcal{T}M_x$ with $M_x$
reflecting $k_x \rightarrow -k_x$, satisfying
\begin{eqnarray}
H(k_x,k_y)=H^*(-k_x,\,k_y),
\end{eqnarray}
which guarantees that the energy spectrum remains even in $k_x$.

Consequently, $m_z$ serves as a single tuning parameter for the system's symmetry.
The chiral limit ($m_z = 0$) suppresses Berry-curvature channels while preserving the quantum metric,
whereas a finite $m_z$ reactivates them.

\subsection{Quantum geometric tensor}

\begin{figure*}
\centering
\includegraphics[width=\linewidth]{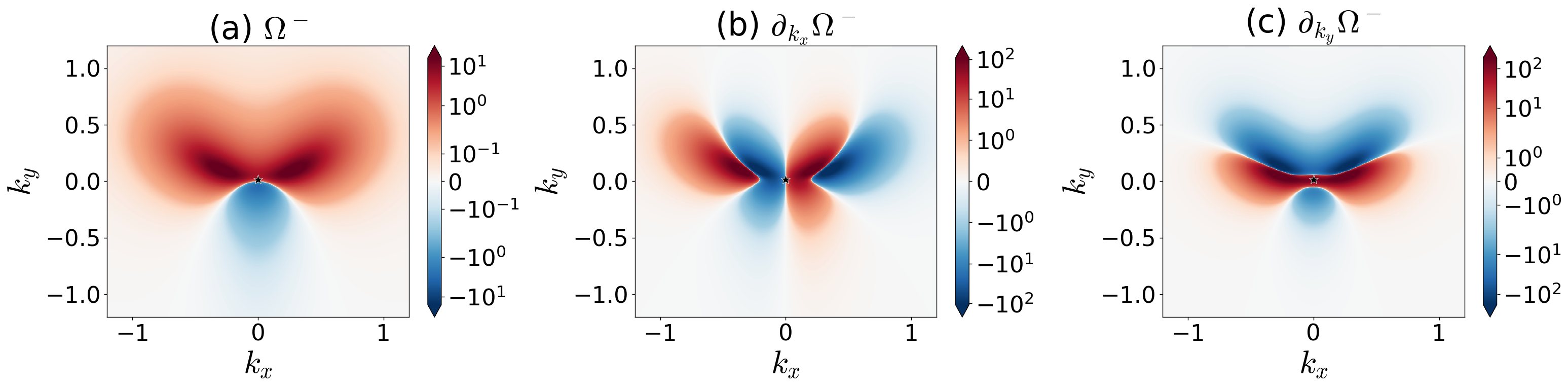}
\\ \vspace{0.3cm}
\includegraphics[width=\linewidth]{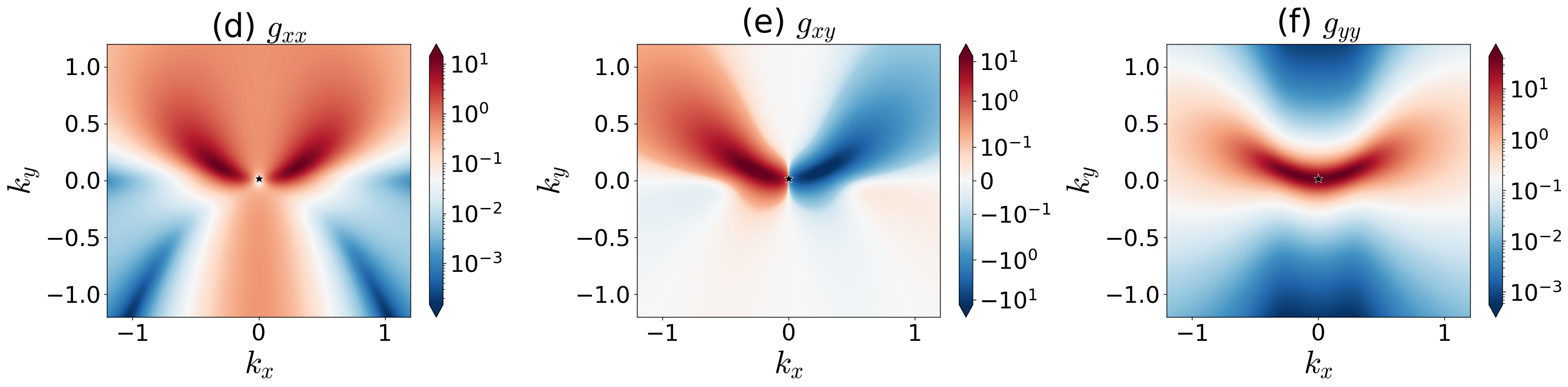}
\vspace{-0.15cm}
\caption{(Color online) Momentum-space distributions for (a) the BC $\Omega^-$, (b)-(c) the density of BC dipole (BCD)
components $\partial_{k_x}\Omega^-$ and $\partial_{k_y}\Omega^-$, and (d)-(f) the QM components.
The black star mark the nodal point, and the parameters used are $\gamma_1 = 0.75$, $v = 0.65$,
$\delta_0 = 0.01$, $\gamma_2 = 1.0$, and $m_z = 0.05$.}
\label{fig:BC_QM}
\end{figure*}

Next, we derive the QGT for the type-II SDSM~(\ref{Eq_H2}), where its symmetric and antisymmetric components correspond to the QM and BC, respectively. By defining the unit vector $\mathbf{n}\equiv\mathbf{h}(\mathbf{k})/|\mathbf{h}(\mathbf{k})|$,
the QGT is expressed as follows~\cite{Ezawa2024PRB,Wiedmann2025PRB,Liu2025NSR,Li2025arXiv,Bera2026arXiv,Villegas2025PRB},
\begin{eqnarray}
G_{\mu\nu}^\xi
&=&g_{\mu\nu}^\xi-\frac{i}{2}\Omega_{\mu\nu}^\xi,\label{eq:QG}
\end{eqnarray}
with $\xi = \pm 1$ denoting the two bands $E^\pm$. Here, $g_{\mu\nu}^\xi$ and $\Omega_{\mu\nu}^\xi$ are designated by~\cite{Ezawa2024PRB,Wiedmann2025PRB,Liu2025NSR,Li2025arXiv,Bera2026arXiv,Villegas2025PRB}
\begin{eqnarray}
g_{\mu\nu}^\xi
&\equiv&\frac{1}{4}\partial_\mu\mathbf{n}\cdot\partial_\nu\mathbf{n},\hspace{0.35cm}
\Omega_{\mu\nu}^\xi\equiv\frac{1}{2}\mathbf{n}
\cdot(\partial_\mu\mathbf{n}\times\partial_\nu\mathbf{n}).\label{Eq_g_Omega}
\end{eqnarray}

Specifically, the real part of QGT $g_{\mu\nu}^\xi$ represents the QM and describes the quantum distance
$ds^2=g_{\mu\nu}\delta k_\mu \delta k_\nu$, whereas its imaginary part $\Omega_{\mu\nu}^\xi$
denotes the BC and characterizes the local band curvature~\cite{Matsuura2010PRB,Provost1980CMP,
Berry1984PRSA,Souza2000PRB,Resta2011EPJB}. In particular, they satisfy the following inequalities~\cite{Liu2025NSR,Yu2025npjQM,Roy2014PRB,Jackson2015NC,Wang2021PRL}:
\begin{eqnarray}
{\rm Tr}\,g^\xi\geq |\Omega^\xi|,\hspace{0.35cm}
{\rm Det}\,g^\xi\geq\frac{|\Omega^\xi|^2}{4},
\end{eqnarray}
where ${\rm Tr}\,g$ and ${\rm Det}\,g$ refer to the trace and determinant of the matrix $g_{\mu\nu}$, respectively.

For the type-II SDSM, combining Eqs.~(\ref{Eq_H2}) and (\ref{Eq_g_Omega}) yields the BC,
\begin{eqnarray}
\Omega_z^\pm(\mathbf{k})
=\mp\frac{m_z\gamma_2(2\gamma_1k_x^2+vk_y)}{2|\mathbf{h}(\mathbf{k})|^3}.\label{Eq_BC_expression}
\end{eqnarray}
Because $\Omega_z^\pm$ is an even function of $k_x$, it generates a nonzero total Berry flux
and the nontrivial density of BC dipole (BCD) components $\partial_{k_x,k_y}\Omega^-$ as shown in
Fig.~\ref{fig:BC_QM}(a)-(c), leading to a nonzero Chern number
$C=\pm 2{\rm sgn}\,(m_z)$~\cite{Huang2015PRB,Bian2026arXiv}. The prefactor of 2 arises from
the four symmetry-related semi-Dirac points in the extended Brillouin zone
(BZ)~\cite{Huang2015PRB,Bian2026arXiv}. This clearly indicates that when the gap closes
at $m_z=0$, the $\mathcal{PT}$ symmetry forces the BC to vanish, eliminating its contribution
to the topological response.

Then, we move to the QM. Since the conduction and valence bands possess the same QM,
its properties are fully captured by analyzing
a single band. For simplicity, we therefore drop the band index $\xi$ from now on.
Following a parallel procedure, we substitute Eqs.~(\ref{Eq_H2}) and (\ref{Eq_g_Omega})
to derive the symmetric QM matrix elements for the 2D SDSM,
\begin{eqnarray}
g_{xx}
&=&
\frac{4\gamma_1^2 k_x^2 + \gamma_2^2 k_y^2 - (\partial_x h)^2}{4h^2}, \label{Eq_g_xx}\\
g_{yy}
&=&
\frac{v^2 + \gamma_2^2 k_x^2 - (\partial_y h)^2}{4h^2},\\
g_{xy}
&=&
\frac{-2\gamma_1 v k_x + \gamma_2^2 k_x k_y - (\partial_x h)(\partial_y h)}{4h^2}=g_{yx},
\end{eqnarray}
where we define $h\equiv|\mathbf{h}(\mathbf{k})|$ and the derivatives $\partial_{x,y} h$ are given by
\begin{eqnarray}
\hspace{-0.3cm}
\partial_x h
=\!\frac{2\gamma_1 k_x h_1 + \gamma_2 k_y h_2}{h},\,\,
\partial_y h
=\!\frac{-v h_1 + \gamma_2 k_x h_2}{h}.\label{Eq_partial_x_y}
\end{eqnarray}

The momentum-space distributions of these matrix elements are depicted in panels (d)-(f) of Fig.~\ref{fig:BC_QM}.
Because of the intrinsic mirror symmetry of the system, the diagonal components $g_{xx}$ and $g_{yy}$
are even functions of $k_x$, whereas the off-diagonal component $g_{xy}$ is an odd function.
This distinct parity behavior is physically significant, as the even diagonal elements yield a finite
net response, whereas the odd-parity off-diagonal terms cancel out upon integration over the entire
BZ. Consequently, the net first-order QM-dependent response vanishes~\cite{Li2026PRL}, which will be explored in detail
in the forthcoming section~\ref{SecIII}.

\begin{figure}
\centering
\includegraphics[width=\linewidth]{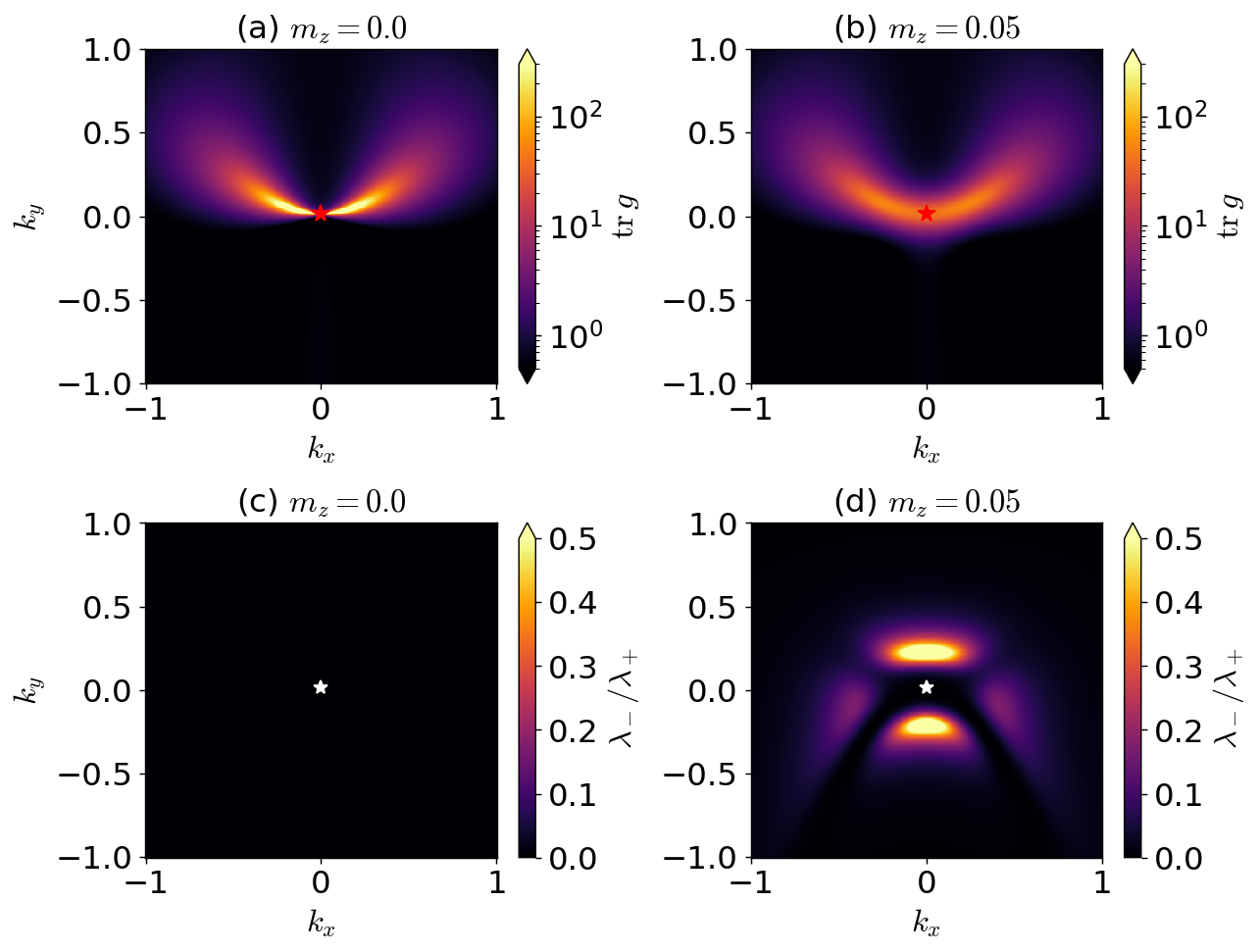}
\vspace{-0.5cm}
\caption{(Color online) Momentum-space distributions of the quantum metric trace $\mathrm{tr}\,g$,
and the anisotropy ratio $\lambda_-/\lambda_+$ for $m_z=0$ (a)-(b) and $m_z=0.05$ (c)-(d).
The star marks the nodal point, and the relevant parameters are chosen as $\gamma_1 = 0.75$,
$v = 0.65$, $\delta_0 = 0.01$, and $\gamma_2 = 1.0$.}
\label{fig:QM}
\end{figure}

\subsection{Quantum metric singularity and tunable isolation}\label{Sec_IID}

Based on the QM expressions in Eqs.~(\ref{Eq_g_xx})-(\ref{Eq_partial_x_y}), we note that $g_{\mu\nu}$
is not directly proportional to $m_z$, allowing the QM to maintain a finite value as the chiral
limit ($m_z \rightarrow 0$) is approached. In sharp contrast, Eq.~(\ref{Eq_BC_expression}) indicates
that the BC exactly vanishes at this point. This implies that $m_z$ not only determines the minimum band
gap $2|m_z|$ but also serves as the key parameter for tuning the interplay between the QM and the BC.

Using Eqs.~(\ref{Eq_g_xx})-(\ref{Eq_partial_x_y}), the QM matrix $g$ is written as
\begin{eqnarray}
g=
\begin{pmatrix}
g_{xx} & g_{xy} \\
g_{yx} & g_{yy}
\end{pmatrix}.\label{Eq_QM_matrix}
\end{eqnarray}
Its eigenvalues, denoted by $\lambda_+$ and $\lambda_-$, which encode the principal
quantum information, can be calculated numerically. Physically, the trace reflects the average
quantum distance density of infinitesimal momentum-space elements, while the difference between the
two eigenvalues characterizes the degree of anisotropy in the QM. We therefore focus on two key quantities,
namely the trace $\mathrm{tr}\,g\equiv\lambda_++\lambda_-$ and
the anisotropy ratio $\lambda_-/\lambda_+$~\cite{Biscocho2025PRB,Li2026arXiv}, and investigate
their dependence on the parameter $m_z$.

\begin{figure}
\centering
\includegraphics[width=\linewidth]{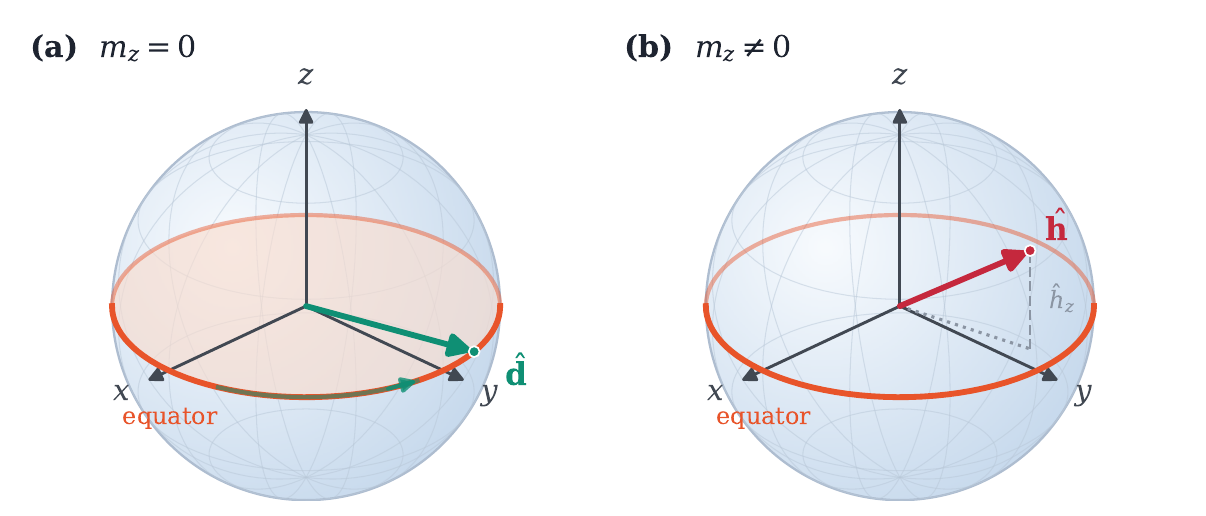}
\vspace{-0.5cm}
\caption{(Color online) Illustration of the unit vector on the Bloch sphere for
(a) $m_z = 0$ ($\hat{\mathbf{h}}=\hat{\mathbf{d}}$) and (b) $m_z \neq 0$ ($\hat{\mathbf{h}}$).
The vector is confined to the equator in the chiral limit (a),
corresponding to a purely real QM, while a finite gap (b) induces a $z$-component that generates the BC.}
\label{fig_schematic_diagram}
\end{figure}

In the chiral limit ($m_z=0$), Figs.~\ref{fig:QM}(a) and (c)
show that ${\rm Tr}\,g$ diverges at the node while $\lambda_-/\lambda_+$ vanishes.
As schematically depicted in Fig.~\ref{fig_schematic_diagram}(a), this indicates
that the QM degenerates to rank 1 and the vector $\mathbf{h}(\mathbf{k})$
lies on the equator of the Bloch sphere.
As a result, the unit vector of $\mathbf{h}(\mathbf{k})$~(\ref{Eq_H2})
is entirely determined by its azimuthal angle,
\begin{eqnarray}
\hat{\mathbf{h}}\equiv\frac{\mathbf{h}(\mathbf{k})}{|\mathbf{h}(\mathbf{k})|}
=\hat{\mathbf{d}}\equiv\frac{\mathbf{d}(\mathbf{k})}{|\mathbf{d}(\mathbf{k})|}
=(\cos\varphi,\sin\varphi,0),
\end{eqnarray}
where $\varphi\equiv {\rm arg}\, (h_1+ih_2)$. The QM can then be simply expressed
as $g_{\mu\nu}=\partial_\mu\varphi\partial_\nu\varphi/4$. Physically, this implies that
only momentum perturbations parallel to the gradient $\nabla\varphi$ (i.e., $d\mathbf{k} \parallel \nabla\varphi$)
alter the Bloch states~\cite{Kim2025SCI,Mera-Goldman2022SPP}, whereas perpendicular perturbations leave them invariant.
Consequently, the QM captures only the longitudinal geometric information of the energy bands.
Because the BC vanishes completely in this limit, the symmetric (QM) and antisymmetric (BC) parts
of the QMT are cleanly separated at $m_z=0$.

\begin{figure*}
\centering
\subfigure[]
{\includegraphics[width=2.25in]{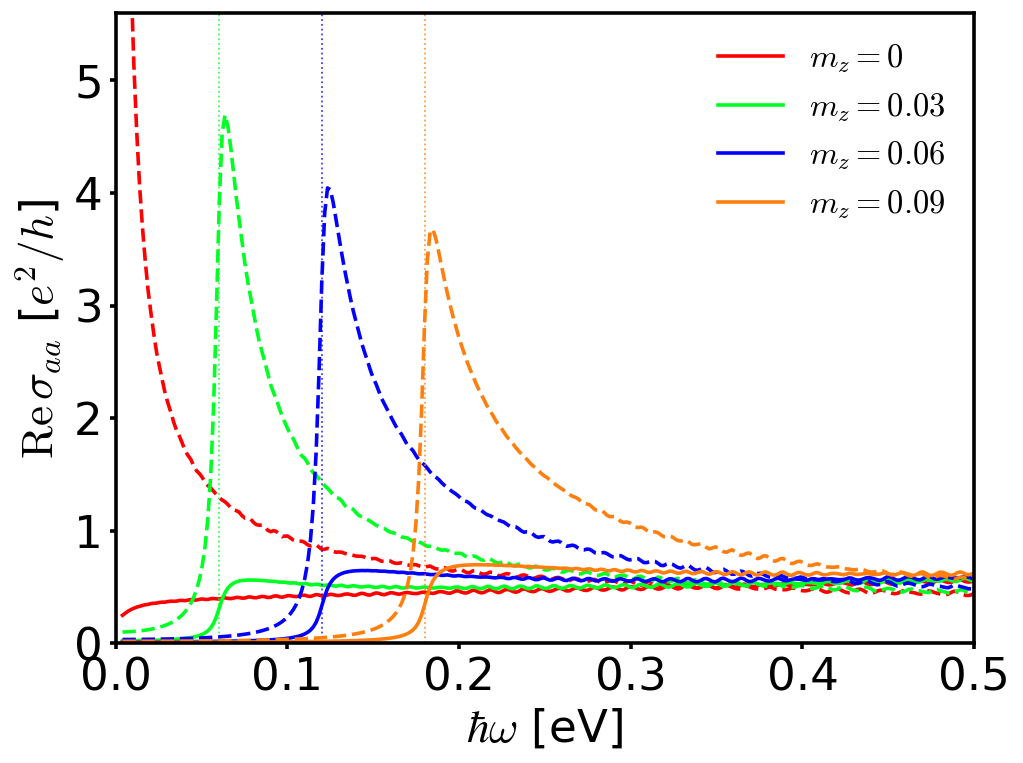}}\hspace{-0.05cm}
\subfigure[]
{\includegraphics[width=2.25in]{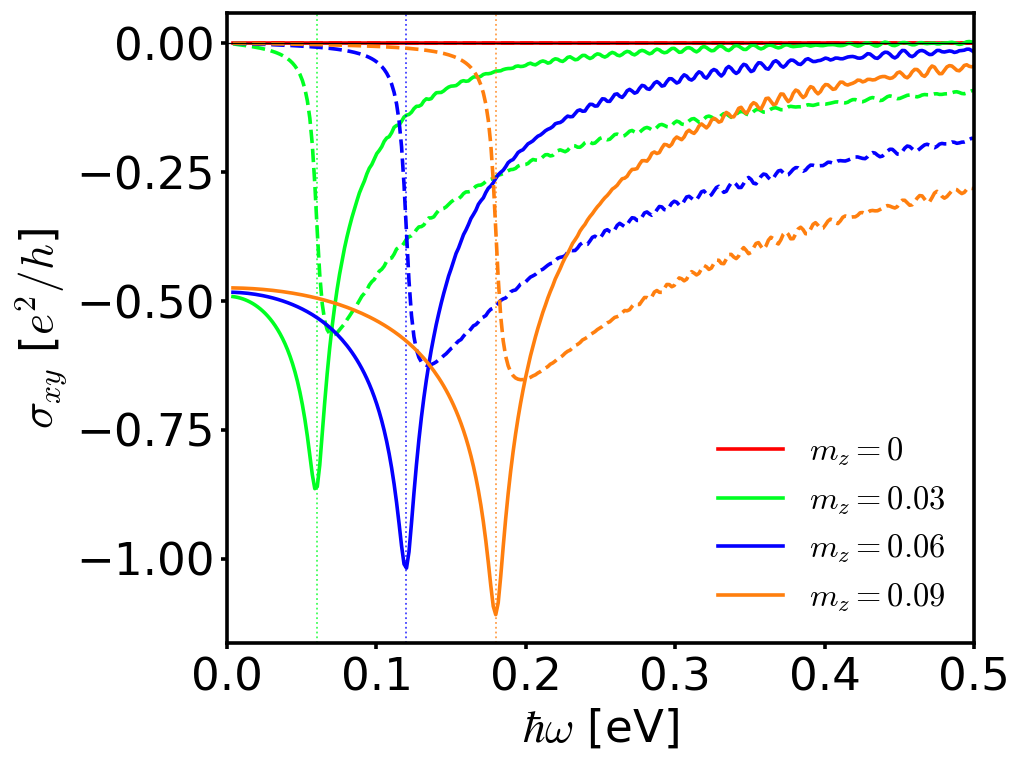}}\hspace{-0.05cm}
\subfigure[]
{\includegraphics[width=2.25in]{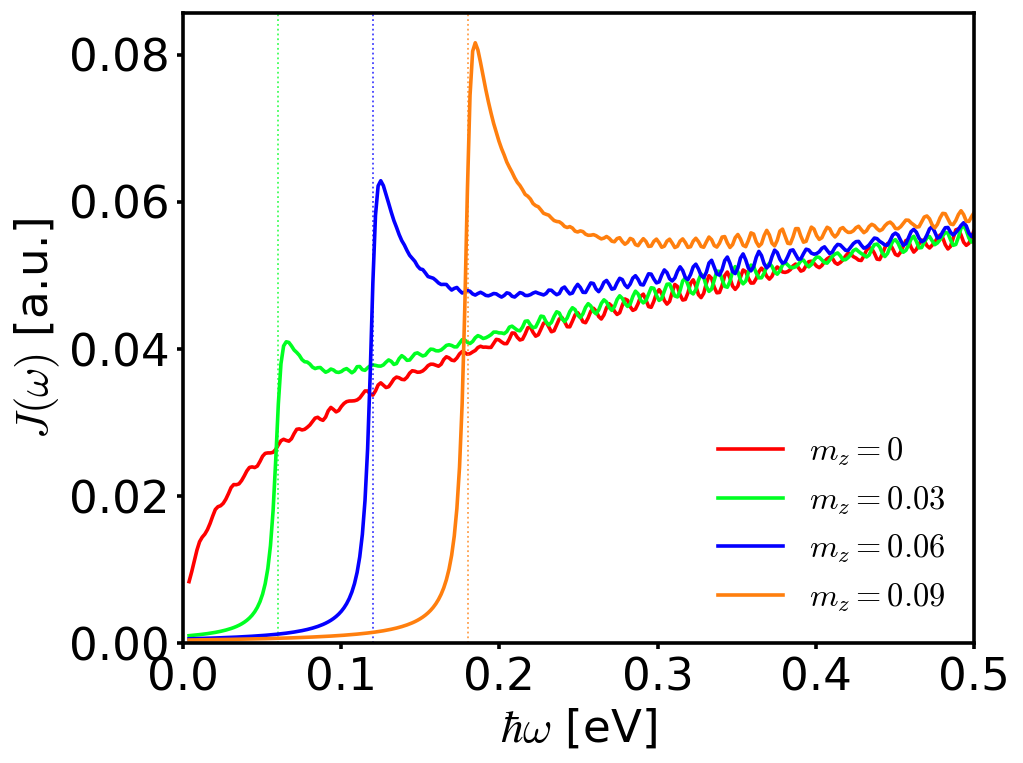}}
\vspace{-0.2cm}
\caption{(Color online) Interband optical properties for different $m_z$ values: (a) Re $\sigma_{xx}$ (solid lines)
and Re $\sigma_{yy}$ (dashed lines), (b) Re $\sigma_{xy}$ (solid lines) and Im $\sigma_{xy}$ (dashed lines),
and (c) JDOS $J(\omega)$.  Vertical dotted lines in (a) and (b) indicate the absorption thresholds at
$\Delta_{\min} = 2|m_z|$ and the relevant parameters are chosen as $\gamma_1 = 0.75$, $v = 0.65$,
$\delta_0 = 0.01$, and $\gamma_2 = 1.0$.}
\label{fig:optical_conduc}
\end{figure*}

When a finite $m_z$ is introduced, as depicted in Figs.~\ref{fig:QM}(b) and (d),
${\rm Tr}\, g$ decreases while $\lambda_-/\lambda_+$ increases.
This evolution reflects the fact that the gap opening suppresses the
singular divergence at the node and restores the second eigenvalue of the QM tensor.
As a result, the unit vector $\hat{\mathbf{h}}$ departs from the equator of the Bloch sphere,
lifting the rank-1 degeneracy of the QM. From a physical standpoint, as the vector moves into
the northern or southern hemisphere, the non-zero $z$-component of $\hat{\mathbf{h}}$
activates a finite BC~\cite{Kolodrubetz2017PR,Fuchs-Montambaux2010EPJB,Peotta2025book} as shown in Fig.~\ref{fig_schematic_diagram}(b).
Therefore, the system is no longer governed by the QM alone, instead, the QG transitions from being purely symmetric
(QM-dominated) to a mixed state where both QM and BC coexist and jointly shape the physical responses.

To wrap up, in Type-II SDSMs, approaching the chiral limit ($m_z \rightarrow 0$) closes the BC channel
and leaves a singular QM, which strongly modulates the Bloch states.
This provides a tunable platform for probing intrinsic geometric effects.
In the following sections, we demonstrate how this isolated QM singularity
dictates the optical and transport signatures.

\section{Optical responses}\label{SecIII}

To reveal the intrinsic geometric properties of the energy bands,
we within this section investigate interband photocurrents at different perturbative orders,
including linear photoconductivity and the nonlinear shift current~\cite{Xiao2010RMP,Kubo1957JPSJ,
Ahn2022NP,Sipe2000PRB,Aversa1995PRB,Young2012PRL,Cook2017NC,Resta2024PRL}.

\subsection{Linear optical conductivity}\label{SecIIIA}

We begin by investigating the optical conductivity.
This linear response relates the first-order current density $J_a^{(1)}$ to the applied electric field $\mathcal{E}_b$
via $J_a^{(1)} = \sigma_{ab}^{(1)} \mathcal{E}_b$ with $a,b=x,y$, where $\sigma_{ab}^{(1)}$ denotes the linear optical conductivity tensor~\cite{Dressel2002book,Giuliani2005book,Neto2009RMP}. We then focus specifically on how this response is modulated by the gap parameter $m_z$.

Using the Kubo formula~\cite{Mahan2000book,Altland2010book},
the interband linear optical conductivity for a 2D system is generally expressed as~~\cite{Ezawa2024PRB2,Wiedmann2025PRB,Liu2025NSR,Li2025arXiv,Bera2026arXiv,Kubo1957JPSJ,
Ahn2022NP,Chiu2026PRB,Ghosh2024SA,Oh2016SA,Shah2026arXiv}
\begin{eqnarray}
\sigma_{ab}=\frac{e^2}{\hbar}\int_{\rm BZ}\sum_{n\neq m}f_{nm}\frac{iE_{mn}(\mathbf{k})A_{nm}^a(\mathbf{k})A_{mn}^b(\mathbf{k})}
{E_{nm}(\mathbf{k})+\hbar\omega+i\eta},\label{eq:sigma}
\end{eqnarray}
where $\int_{\rm BZ}\equiv \int d^2\mathbf{k}/(2\pi)^2$
and $\mathcal{F}_{nm}$ is designated by,
\begin{eqnarray}
\mathcal{F}_{nm}\equiv f(E_n)-f(E_m),\label{Eq_F_mn}
\end{eqnarray}
with
\begin{eqnarray}
f(E)=1/[1+e^{(E-\mu)/k_B T}],\label{Eq_f}
\end{eqnarray}
being the Fermi distribution. Here, $E_n$ denotes the energy eigenvalue of the $n$-th band, $E_{nm} = E_n - E_m$, $\mu$ is the
chemical potential, $\hbar\omega$ is the photon energy, and $\eta$ is an infinitesimal broadening parameter.
In particular, $A_{nm}^a(\mathbf{k})$ denotes the interband Berry
connection for $m \neq n$, defined as~\cite{Oh2016SA}
\begin{eqnarray}
A_{nm}^a(\mathbf{k})=i\langle u_n(\mathbf{k})|\partial_{k_a}|u_m(\mathbf{k})\rangle.
\end{eqnarray}
where $|u_n(\mathbf{k})\rangle$ and $|u_m(\mathbf{k})\rangle$ are the eigenstates
of the Hamiltonian for the $n$-th and $m$-th bands, respectively.

Using this definition, the quantum metric in Eq.~(\ref{Eq_g_Omega}) can be
cast into an equivalent form~\cite{Liu2025NSR},
\begin{eqnarray}
g_{ab}={\rm Re}\sum_{n\neq m}\frac{\langle u_n|\partial_{k_a}H|u_m\rangle\langle u_m|\partial_{k_b}H|u_n\rangle}
{E_{nm}^2}.\label{Eq_g_aa_v2}
\end{eqnarray}
For the two-band system, we label the bands as $m,n = \pm$. By taking the real part of
the diagonal elements in Eq.~(\ref{eq:sigma}) and substituting the diagonal elements of QM ($g_{aa}$) from Eq.~(\ref{Eq_g_aa_v2}),
the longitudinal linear optical conductivity for a 2D system is obtained as~\cite{Bera2026arXiv,Wiedmann2025PRB,Kubo1957JPSJ,Chiu2026PRB,Ghosh2024SA,Ezawa2024PRB2}
\begin{eqnarray}
{\rm Re}\,\sigma_{aa}=\frac{\pi e^2}{\hbar}\int_{\rm BZ}\mathcal{F}_{-+}\Delta \delta(\hbar\omega-\Delta)g_{aa},
\end{eqnarray}
where $\Delta\equiv E_+-E_-$, and $a=x,y$. The delta function indicates that interband optical
absorption occurs only when the photon energy exceeds the band gap ($\hbar\omega > \Delta$).

In comparison, by utilizing the antisymmetric elements
of the QGT given in Eq.~(\ref{eq:QG}), we obtain the transverse optical Hall conductivity, which is determined by the BC~\cite{Bera2026arXiv,Wiedmann2025PRB,Kubo1957JPSJ,Chiu2026PRB,Ghosh2024SA,Ezawa2024PRB2,Li2025arXiv,Ahn2022NP,Oh2016SA,Shah2026arXiv},
\begin{eqnarray}
\hspace{-0.55cm}
{\rm Re}\,\sigma_{xy}&=&\frac{-2e^2}{\hbar}\!\!\int_{\rm BZ}\!\!\frac{\mathcal{F}_{-+}|\mathbf{h}|^2}{\Delta}\mathcal{P}\!\left(\frac{2\Delta}{\Delta^2-\omega^2}\right)\!\Omega_z(\mathbf{k}),\\
{\rm Im}\,\sigma_{xy}
&=&\frac{-2\pi e^2}{\hbar}\!\!\int_{\rm BZ}\!\!\mathcal{F}_{-+}|\mathbf{h}|^2\delta(\hbar\omega-\Delta)\Omega_z(\mathbf{k}),
\end{eqnarray}
where $\mathcal{P}$ serves as the Cauchy principal value. Notably, in the static limit ($\omega \rightarrow 0$),
its real component reduces to the static anomalous Hall conductivity (AHC)~\cite{Okamoto2026CP}, ${\rm Re}\,\sigma_{xy}(0)=-\frac{e^2}{\hbar}\int_{\mathrm{BZ}}\Omega_z=-C\cdot\frac{e^2}{\hbar}$ with $C$
the Chern number. Accordingly, the symmetric component (QM) of the QGT is responsible for the
longitudinal linear optical conductivity, whereas the antisymmetric component (BC) dictates the
transverse optical Hall conductivity.

These distinct geometric contributions are clearly resolved in the numerically calculated
spectra shown in Fig.~\ref{fig:optical_conduc}. Panels (a) and (b) show the QM-governed
optical conductance and the BC-driven optical Hall conductivity for the type-II 2D SDSM, respectively.
In the chiral limit ($m_z=0$), the BC-induced optical response vanishes,
and only the QM channel remains finite. Particularly, ${\rm Re}\, \sigma_{yy}$
diverges as $\hbar\omega \to 0$, reflecting the singular geometry discussed in Sec.~\ref{Sec_IID}.
For a finite $m_z$, the optical conductance exhibits a distinct peak, with its peak energy
$\hbar\omega_p$ shifting to higher values as $m_z$ increases. A similar single-peak phenomenon
is also observed in the Hall response shown in Fig.~\ref{fig:optical_conduc}(b). Notably,
as $\hbar\omega \rightarrow 0$, ${\rm Re}\,\sigma_{xy}$ reduces to the static AHC for finite $m_z$,
consistent with ${\rm Re}\,\sigma_{xy}(0)=-C\cdot\frac{e^2}{\hbar}$, where the half-integer
conductance of $1/2\,(e^2/\hbar)$ from a single Dirac cone sums up to the total Chern number $C$.

To clarify the origin of these peaks, let us consider the joint density of states (JDOS)~\cite{Qian2023npjCM,Yoshida2025PRB},
\begin{eqnarray}
J(\omega)\equiv\int_{\mathrm{BZ}}\delta(\hbar\omega-\Delta(\mathbf{k})),
\end{eqnarray}
which quantifies the density of available states for vertical interband transitions at a given photon energy~\cite{Gerbi2025AIPA,Gu2023N,Nielsen2023arXiv}. Figure~\ref{fig:optical_conduc}(c) shows the
evolution of the JDOS as a function of $\hbar\omega$ for different values of $m_z$ used in panels (a) and (b). Clearly, a
distinct peak appears at $\hbar\omega = 2m_z$. This occurs because $2m_z$ corresponds to
the minimum direct gap, at which the divergent density of states near the nodal point
leads to a pronounced JDOS peak when the photon energy matches this gap.
Table~\ref{tab:mz_dependence} confirms the scaling of the peak position $\hbar\omega_p$ with $2m_z$.
As a result, the optical conductivity peaks exactly at this JDOS maximum.

\begin{table}
\centering
\caption{Evolution of the minimum band gap $\Delta_{\min}$ and the peak energies $\hbar\omega_p$
 for the four optical response channels as a function of $2m_z$.}
\setlength{\tabcolsep}{12pt} 
\vspace{0.2cm}
\begin{tabular}{l|cccc}
\hline
\hline
$2m_z$ & 0.00 & 0.06 & 0.12 & 0.18 \\
\hline
$\Delta_{\rm min}$ & 0.00 & 0.06 & 0.12 & 0.18 \\ 
$\hbar\omega_p^{{\rm Re}\sigma_{xx}}$ & 0.004 & 0.078 & 0.146 & 0.214\\
$\hbar\omega_p^{{\rm Re}\sigma_{yy}}$ & 0.004 & 0.063 & 0.123 & 0.184 \\
$\hbar\omega_p^{{\rm Re}\sigma_{xy}}$ & 0.004 & 0.058 & 0.120 & 0.179 \\
$\hbar\omega_p^{{\rm Im}\sigma_{xy}}$ & 0.004 & 0.070 & 0.135 & 0.196 \\
\hline
\hline
\end{tabular}
\label{tab:mz_dependence}
\end{table}

To be brief, while the QM exclusively dictates the optical response in the chiral limit ($m_z=0$),
a finite $m_z$ activates a distinct peak arising from the geometric band structure and manifested in the JDOS.
These features serve as clear measurable optical fingerprints of the underlying QG.

\begin{figure}
\centering
\subfigure[]
{\includegraphics[width=0.8\linewidth]{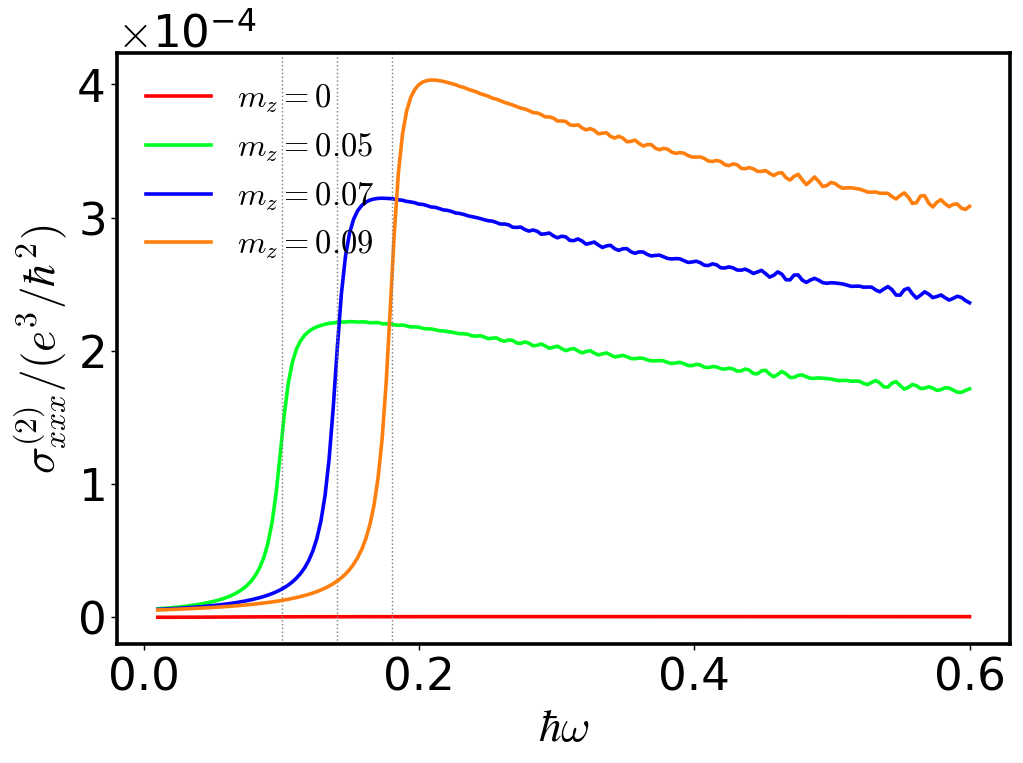}}
\subfigure[]
{\includegraphics[width=0.8\linewidth]{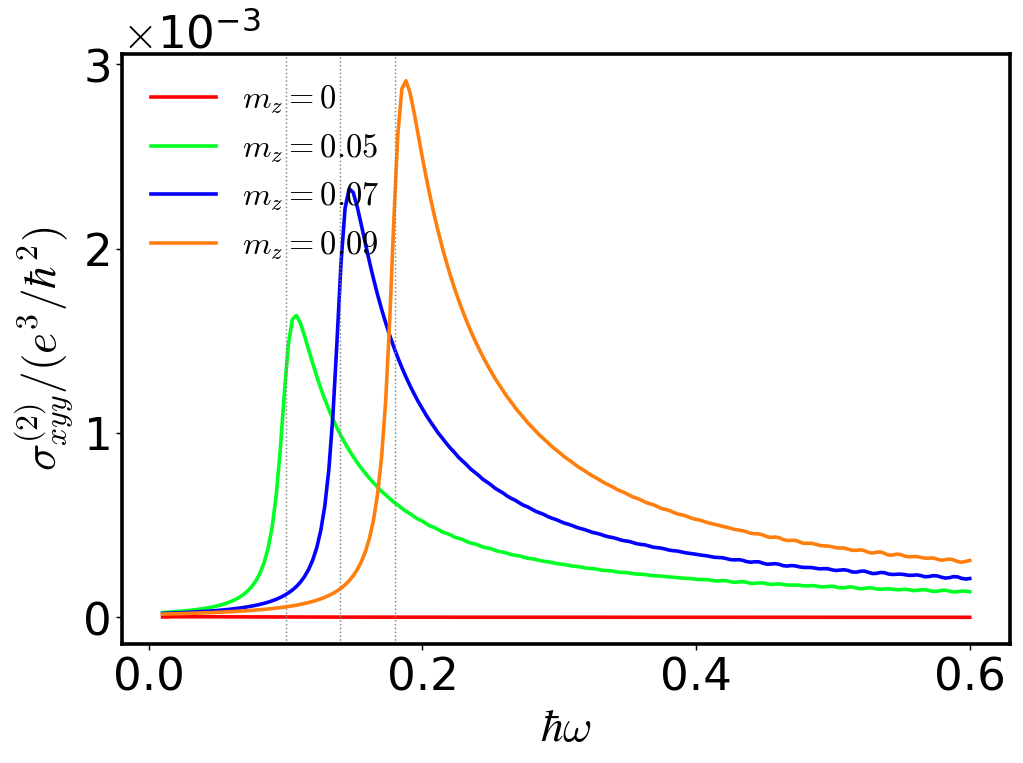}}
\vspace{-0.2cm}
\caption{(Color online) Shift-current conductivity (a) $\sigma_{xxx}^{(2)}$ and (b) $\sigma_{xyy}^{(2)}$
(in units of $e^3/\hbar^2$) as a function of photon energy $\hbar\omega$ for different values of
$m_z$. Vertical dotted lines mark the thresholds $2|m_z|$. The parameters used are $\gamma_1 = 0.75$,
$\gamma_2 = 1.0$, $v = 0.65$, and $\delta_0 = 0.01$.}
\label{fig:shiftcurrent}
\end{figure}

\subsection{Nonlinear shift current}\label{SecIIIB}

Next, we evaluate the shift current response of the type-II SDSM. This second-order optical
response is given by $J_b = \sigma_{bac} \mathcal{E}^a \mathcal{E}^c$, where $J_b$ denotes the photocurrent density,
$\sigma_{bac}$ is the second-order nonlinear shift current conductivity tensor, and $\mathcal{E}^{a,c}$
are the components of the applied electric field~\cite{Qian2023npjCM,Yoshida2025PRB,Baltz1981PRB,Sipe2000PRB,
Yan2018arXiv,Tan2016npjCM,Rangel2017PRL,Resta2024PRL}.

When the type-II SDSM is subjected to an incident light field of frequency $\omega$
linearly polarized along the $a$-direction, it generates a shift-current surface conductivity
along the $b$-direction. This response is intrinsically governed by the difference in Berry
connections between the valence and conduction bands and is expressed as~\cite{Aversa1995PRB,Young2012PRL,Cook2017NC,Rangel2017PRL,Yan2018arXiv,Resta2024PRL,Wang2025CPL}
\begin{eqnarray}
\sigma_{baa}^{(2)}(\omega)
=\frac{\pi e^3}{\hbar^2}\int_{\mathrm{BZ}}\mathcal{F}_{-+}|M_a|^2R_b\delta(\hbar\omega-\Delta),
\end{eqnarray}
where $\mathcal{F}_{-+}$ is the Fermi distribution difference defined in Eq.~(\ref{Eq_F_mn}),
$\Delta = E_+ - E_-$ with $\pm$ band indexes.
Hereby, $M_a\equiv\langle u_-|\partial_{k_a}H|u_+\rangle$ is the interband transition
matrix element, and $R_b$ is the shift vector, defined by~\cite{Resta2024PRL,Rangel2017PRL}
\begin{eqnarray}
R_b&=&
{\rm Im}\left(\frac{\partial_{k_b}A_{-+}^a}{A_{-+}^a}\right)+A_{++}^b-A_{--}^b,\label{eq:R_b}
\end{eqnarray}
with $A_{-+}^a=M_a/(i\Delta)$. Due to the $\mathcal{T}M_x$ symmetry, the shift vectors
satisfy $R_x(k_x) = R_x(-k_x)$ and $R_y(-k_x) = -R_y(k_x)$. Since the odd function $R_y$
integrates to zero over the symmetric Brillouin zone, only the $b=x$ component yields
a non-zero shift current.

To proceed, let us focus on $m_z=0$.
From Eq.~(\ref{Eq_H}), chiral symmetry dictates $\sigma_z H\sigma_z = -H$.
For a conduction band state $|u_+\rangle$ with $H|u_+\rangle=E_+|u_+\rangle$ ($E_+>0$), we find
\begin{eqnarray}
H(\sigma_z|u_+\rangle)
=-\sigma_z H|u_+\rangle=-E_+(\sigma_z|u_+\rangle).
\end{eqnarray}
This shows that $\sigma_z|u_+\rangle$ is the valence band state $|u_-\rangle$, namely,
$\sigma_z|u_+\rangle=e^{i\varphi(\mathbf{k})}|u_-\rangle$,
where the global phase $\varphi(\mathbf{k})$ reflects the $U(1)$ gauge freedom in the BZ.
The intraband and interband Berry connections thus satisfy
\begin{eqnarray}
A_{--}^b &=& i\langle u_-|\partial_{k_b} u_-\rangle = \partial_{k_b}\varphi+A_{++}^b,\\
A_{-+}^a &=&e^{i\varphi}\langle \sigma_z u_+|i\partial_{k_a} u_+\rangle \equiv e^{i\varphi}K_a,
\end{eqnarray}
where $K_a \equiv \langle\sigma_z u_+|i\partial_{k_a} u_+\rangle=K^*_a$. We then obtain
\begin{equation}
{\rm Im}\left(\frac{\partial_{k_b}A_{-+}^a}{A_{-+}^a}\right)
= {\rm Im}\left(i\partial_{k_b}\varphi+\frac{\partial_{k_b}K_a}{K_a}\right) = \partial_{k_b}\varphi.
\end{equation}
Substituting these into Eq.~(\ref{eq:R_b}), the shift vector in Eq.~(\ref{eq:R_b}) vanishes identically
$R_b = \partial_{k_b}\varphi - \partial_{k_b}\varphi = 0$, yielding
$\sigma_{baa}^{(2)}\big|_{m_z=0}=0$.
Physically, chiral symmetry locks the valence and conduction bands into mirror-image configurations, yielding a strictly vanishing optical shift current.

In comparison, introducing a finite spin-orbit gap $m_z \neq 0$ explicitly breaks the
chiral symmetry ($\{\sigma_z, H\} \neq 0$). The strict mirror-image relation
$\sigma_z|u_+\rangle = e^{i\varphi(\mathbf{k})}|u_-\rangle$ no longer holds,
which lifts the exact cancellation in the shift vector $R_b$.
Consequently, $R_b$ becomes finite, and the shift-current response is turned on.

Based on the above analysis, $m_z$ acts as a highly effective knob for tuning the shift-current conductivity.
Figure~\ref{fig:shiftcurrent} shows the evolution of the shift-current conductivity as a function of photon energy
$\hbar\omega$ for different values of $m_z$. Consistent with our predictions, both $\sigma_{xxx}^{(2)}$
and $\sigma_{xyy}^{(2)}$ vanish entirely in the chiral limit ($m_z=0$). For finite $m_z$, however, a finite
shift current emerges with a distinct peak, whose amplitude and position both increase with $m_z$. These
results confirms that shift current is strictly forbidden by chiral symmetry at $m_z=0$ but becomes highly
tunable once the chiral symmetry is broken by a finite spin-orbit gap $m_z$.

To recapitulate, the optical responses reveal a clear dichotomy. A finite $m_z$ generates
the experimentally observed single-peak features in both the linear optical conductivity and the
shift current. Conversely, the chiral limit ($m_z = 0$) causes both the BC-driven optical Hall
conductance and the shift current to vanish. This leaves the QM-induced optical conductivity as
the only surviving observable. Collectively, these results demonstrate that $m_z$ serves as a
single, tunable parameter to isolate and probe intrinsic
quantum geometric effects.

\begin{figure}
\centering
\hspace{-1cm}
\includegraphics[width=1.7in]{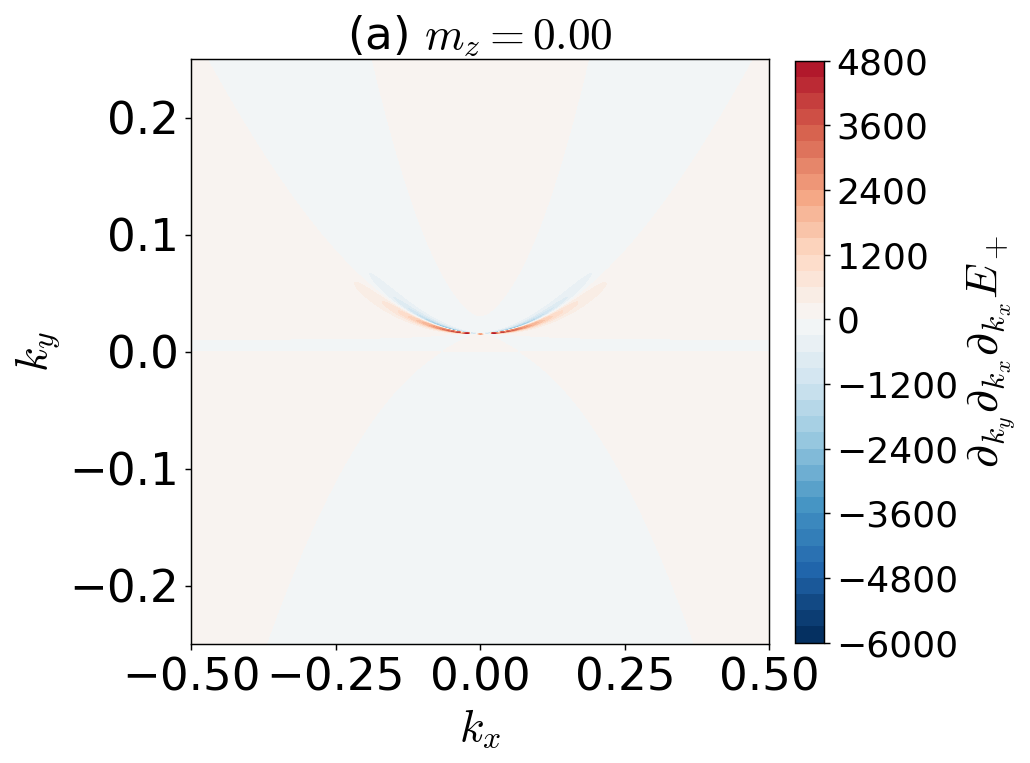}
\includegraphics[width=1.7in]{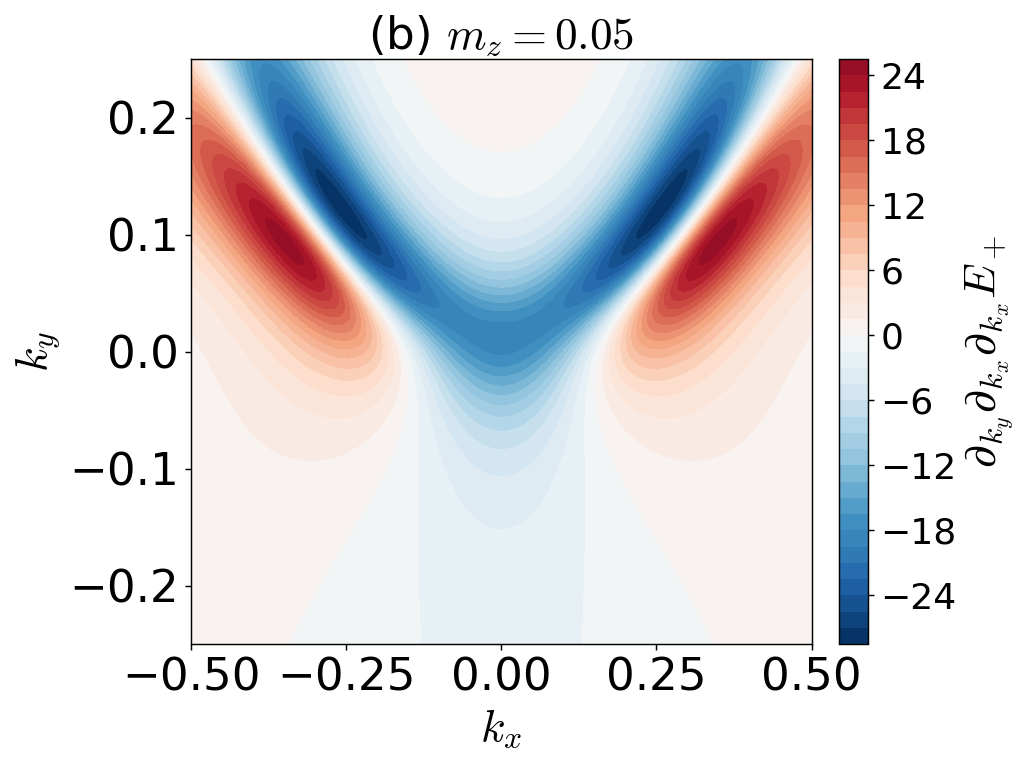}
\vspace{-0.2cm}
\caption{(Color online) Distribution of the Drude conductivity kernel ($\partial_{k_x}^2 \partial_{k_y} E_+$)
in momentum space for (a) $m_z = 0.00$ and (b) $m_z = 0.05$. The parameters are chosen as $\gamma_1 = 0.75$,
$\gamma_2 = 1.0$, $v = 0.65$, and $\delta_0 = 0.01$.}
\label{fig:drude_density}
\end{figure}

While the linear optical responses provide clear signatures of the quantum geometry, nonlinear
transport offers a complementary and equally powerful probe, which will be studied in the following section.


\section{Nonlinear Hall Response and Scaling Analysis}\label{SecIV}

Beyond the optical signatures of the isolated QM, this section examines the nonlinear
Hall effect as a complementary platform to study the Drude and QG contributions.
We also investigate the feasibility of separating
these two components using scaling analysis.
~\cite{Liu2025NSR,Ezawa2024PRB2,Jiang2025RPP,Shibata2026arXiv,Gao2026RMP,Gong2026SB}.

\subsection{Nonlinear Hall conductivity}

The second-order nonlinear conductivity $\sigma_{abc}$ is defined by $j_c = \sigma_{abc} \mathcal{E}^a \mathcal{E}^b$,
where $j_c$ is the current density and $\mathcal{E}^{a,b}$ are the components of the applied electric field~\cite{Liu2025NSR,Ezawa2024PRB2}. Expanding this in terms of the electron relaxation time $\tau$  yields~\cite{Liu2025NSR,Ezawa2024PRB2,Jiang2025RPP,Shibata2026arXiv,Gao2026RMP,Gong2026SB}
\begin{eqnarray}
\sigma_{abc}=\sigma^{\rm Dru}_{abc}+\sigma_{abc}^{\rm BCD}+\sigma_{abc}^{\rm QMD}.
\end{eqnarray}
Here, the Drude term $\sigma^{\rm Dru}_{abc}$ is proportional to $\tau^2$,
while the Berry curvature dipole (BCD) and quantum metric dipole (QMD) terms represent
quantum-geometric contributions scaling as $\tau$ and $\tau^0$, respectively.
We now evaluate each of these contributions for the type-II SDSM.

\begin{figure*}
\centering
\subfigure[]
{\includegraphics[width=2.2in]{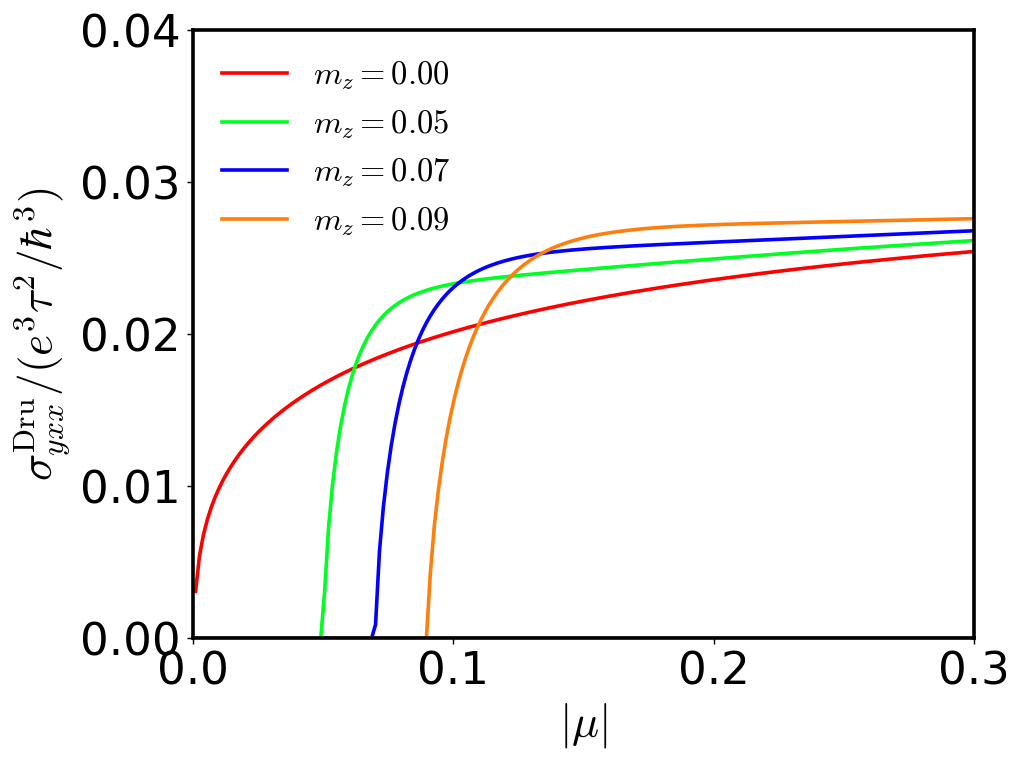}}
\subfigure[]
{\includegraphics[width=2.2in]{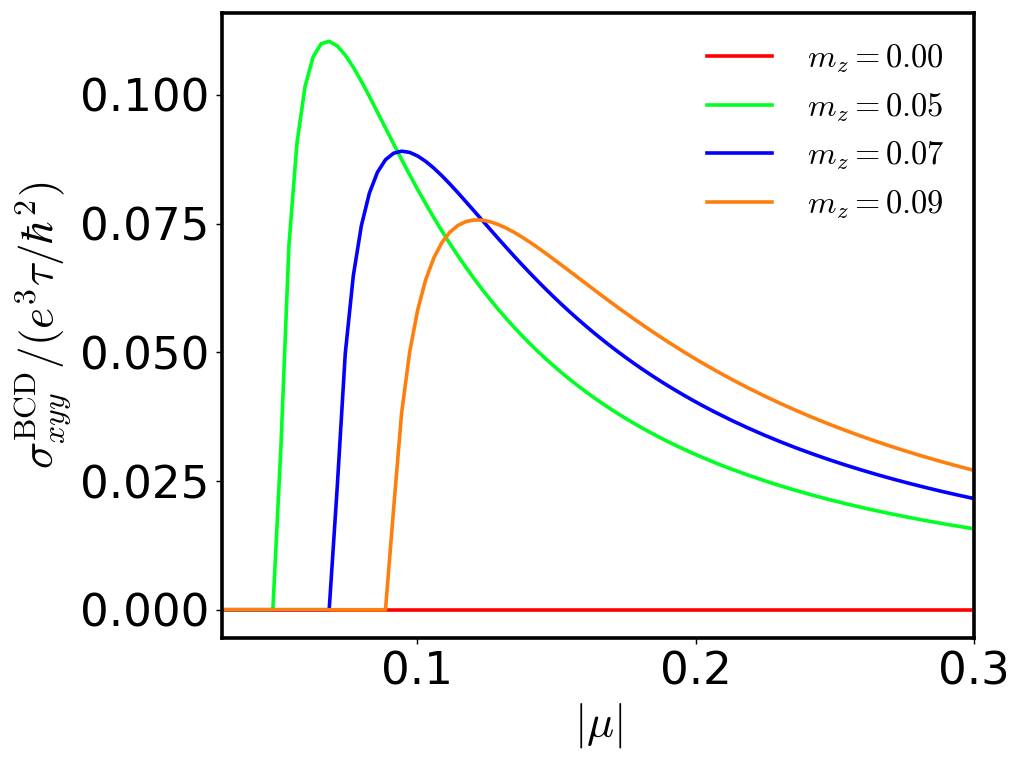}}
\subfigure[]
{\includegraphics[width=2.2in]{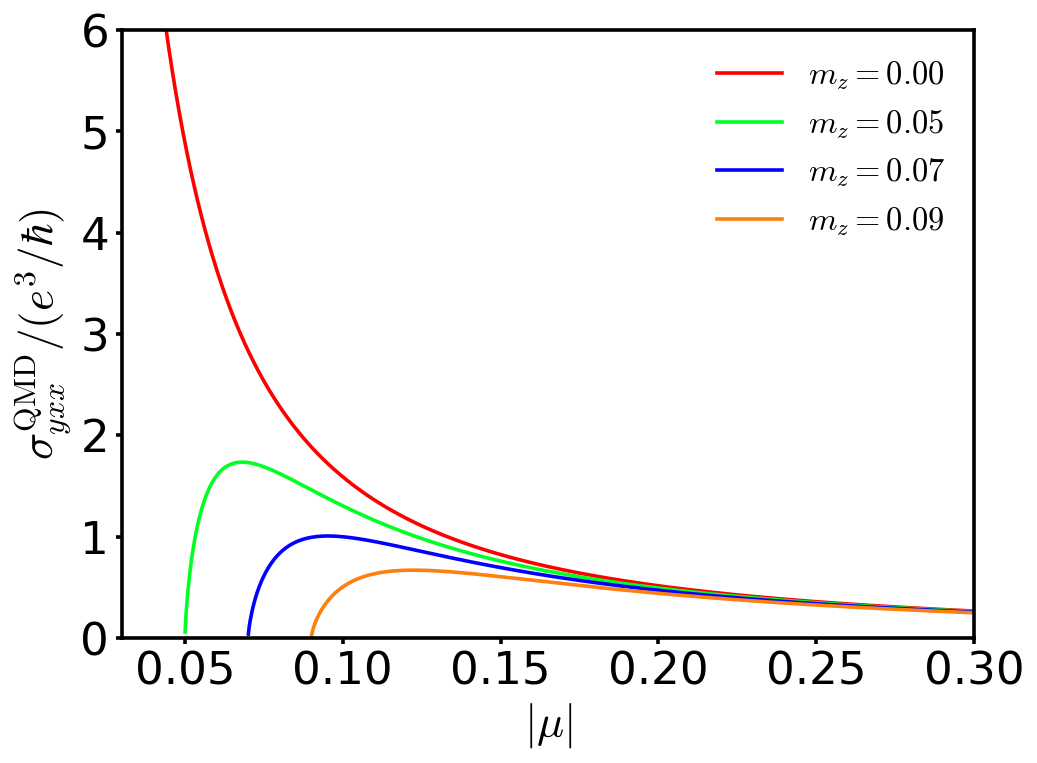}}
\vspace{-0.25cm}
\caption{(Color online) Evolution of the (a) Drude, (b) BCD-induced, and (c) QMD-induced NHC as a function of the chemical potential $|\mu|$ for different values of $m_z$. The vertical axes correspond to the tensor components $\sigma_{yxx}^{\rm Dru}/(e^3\tau^2/\hbar^3)$, $\sigma_{xyy}^{\rm BCD}/(e^3\tau/\hbar^2)$, and $\sigma_{yxx}^{\rm QMD}/(e^3/\hbar)$, respectively. The parameters
are chosen as $\gamma_1 = 0.75$, $\gamma_2 = 1.0$, $v = 0.65$, and $\delta_0 = 0.01$.}
\label{fig:sigma_BCD}
\end{figure*}

We begin with the nonlinear Drude conductivity, which is given by~\cite{Liu2025NSR,Ezawa2024PRB2,Jiang2025RPP,Shibata2026arXiv,Gao2026RMP,Gong2026SB,Kaplan2023SP}
\begin{eqnarray}
\sigma^{\rm Dru}_{abc}=-\frac{e^3\tau^2}{\hbar^3}\sum_n\int_{\mathrm{BZ}}(\partial_{k_a}\partial_{k_b}\partial_{k_c} E_n)f_n,
\end{eqnarray}
which originates from the deformation of the Fermi surface, with the indices $a, b, c \in \{x, y\}$ denoting
the spatial components and $f_n$ defined in Eq.~(\ref{Eq_f}).
Owing to the \(\mathcal{T}M_x\) symmetry as shown in Sec.~\ref{Sec_IIB}, the non-vanishing components of
the nonlinear Drude conductivity are restricted to $\sigma^{\rm Dru}_{yxx}$ and $\sigma^{\rm Dru}_{xxy}$
(with $\sigma^{\rm Dru}_{xxy} = \sigma^{\rm Dru}_{xyx}$), corresponding to the nonlinear currents
along the $x$ and $y$ directions, respectively.

The momentum-space distribution of the Drude kernel $\partial_{k_x}^2 \partial_{k_y} E_+$ is presented in Fig.~\ref{fig:drude_density} (The valence band kernel is identical in structure but opposite in sign,
$\partial_{k_x}^2 \partial_{k_y} E_- = -\partial_{k_x}^2 \partial_{k_y} E_+$).
We notice that it remains finite and is strictly an even function of $k_x$, regardless of whether $m_z$ vanishes or not.
Building on this, Fig.~\ref{fig:sigma_BCD}(a) demonstrates the resulting nonlinear Drude conductivity.
While the Drude conductivity is finite for any value of $m_z$, it vanishes entirely when the chemical potential
lies inside the band gap ($|\mu| < m_z$). This is because the Drude response originates from Fermi surface deformations~\cite{Sodemann2015PRL,Dressel2002AJP}. When the Fermi level lies inside the band gap, there are no active states
to contribute to transport. Additionally, we note that the non-vanishing tensor components are strictly equal,
satisfying $\sigma_{yxx}^{\rm Dru} = \sigma_{xxy}^{\rm Dru} = \sigma_{xyx}^{\rm Dru}$,
which is consistent with the underlying symmetries of the system.

Next, we evaluate the Berry curvature dipole (BCD) contribution to the NHC, which is expressed as~\cite{Liu2025NSR,Ezawa2024PRB2,Jiang2025RPP,Shibata2026arXiv,Gao2026RMP}
\begin{eqnarray}
\sigma_{abc}^{\rm BCD}
=-\frac{e^3\tau}{\hbar^2}\epsilon_{abz}D_c,
\end{eqnarray}
where $\epsilon_{ijk}$ is the Levi-Civita symbol, and the BCD $D_c$ is defined as~\cite{Liu2025NSR,Ezawa2024PRB2,Jiang2025RPP,Shibata2026arXiv,Gao2026RMP,Sodemann2015PRL,Ulrich2026PRB}
\begin{eqnarray}
D_c=\int_{\mathrm{BZ}}\mathcal{F}_{-+}\partial_{k_c}\Omega_z(\mathbf{k}).
\end{eqnarray}
with $\mathcal{F}$ is designated in Eq.~(\ref{Eq_f}).

As shown in Fig.~\ref{fig:BC_QM}(b)-(c) for the BCD density, $\partial_{k_x}\Omega_z$ is odd in $k_x$,
causing $D_x$ to vanish upon BZ integration. In contrast, $\partial_{k_y}\Omega_z$ is even in $k_x$, leaving only $D_y$ finite.
This corresponds to the only non-vanishing component,
\begin{eqnarray}
\sigma_{xyy}^{\rm BCD}=\frac{e^3\tau}{\hbar^2}D_y.
\end{eqnarray}
Figure~\ref{fig:sigma_BCD}(b) presents the evolution of this component of BCD-induced NHC as a
function of chemical potential under varying $m_z$. As anticipated, this channel vanishes at
$m_z=0$ due to the absence of BC. For a finite \(m_z\), the breaking of chiral symmetry activates
the BC, yielding a distinct peak. Importantly, as $m_z$ increases, the widened gap $2|m_z|$ shifts
this peak upward in energy, while the peak amplitude is suppressed, indicating a reduced BC density.

\begin{figure}
\centering
\includegraphics[width=\linewidth]{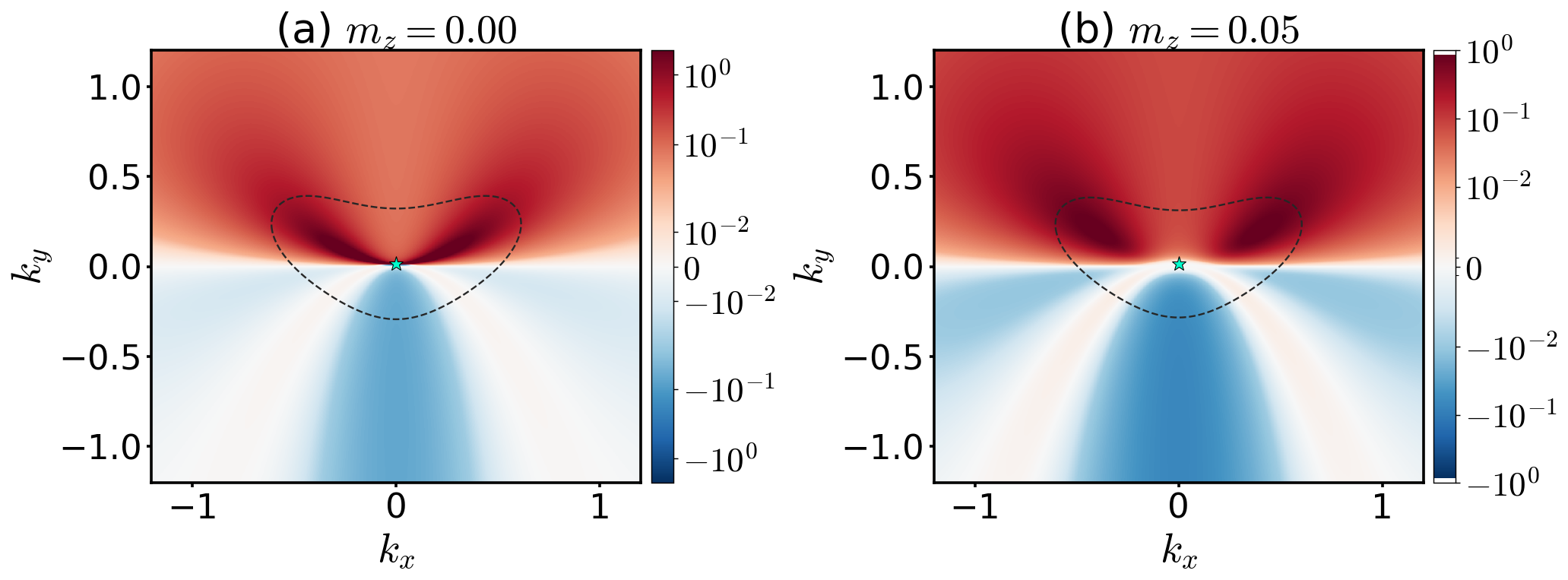}
\vspace{-0.68cm}
\caption{(Color online) Distribution of the QMD kernel
in momentum space for (a) $m_z = 0.00$ and (b) $m_z = 0.05$.
The parameters are chosen as $\gamma_1=0.75$, $\gamma_2=1.0$, $v=0.65$, and $\delta_0=0.01$.}
\label{fig:QMD_kernel}
\end{figure}

Finally, let us consider the QMD-induced NHC. The $\mathcal{T}M_x$ symmetry of the type-II SDSM dictates that only the $yxx$ component survives. The corresponding QMD is defined as~\cite{Bera2026arXiv,Ulrich2026PRB,Ahmad2026PRB},
\begin{eqnarray}
\Lambda_{yxx}^n
&=&\int_{\mathrm{BZ}} (v_y^n g_{xx}^n-v_x^n g_{yx}^n)\delta(E_n-\mu).
\end{eqnarray}
It then gives rise to the nontrivial QMD contribution to the NHC, given by~\cite{Liu2025NSR,Ezawa2024PRB2,Jiang2025RPP,Shibata2026arXiv,Gao2026RMP,Sodemann2015PRL,Ulrich2026PRB}
\begin{eqnarray}
\hspace{-0.15cm}
\sigma_{yxx}^{\rm QMD}
=\frac{-2e^3}{\hbar^2}\sum_n\int_{\mathrm{BZ}}
\frac{(v_y^n g_{xx}^n-v_x^n g_{yx}^n)}{E_n-E_{\bar{n}}}\delta(E_n-\mu).
\end{eqnarray}
Here, $v_x^n$ and $v_y^n$ are the group velocities of the $n$-th band. $E_n$ and $E_{\bar{n}}$
represent the energy dispersions of the $n$-th band and its complementary band ($n \neq \bar{n}$),
respectively. For our two-band model, $n=\pm$ denote the conduction and valence bands, with $\bar{n} = \mp$.

\begin{table}[htbp]
\centering
\caption{Selection rules for the components of the second-order nonlinear conductivity tensor.
The checkmarks ($\checkmark$) and zeros (0) indicate the presence and absence of the corresponding
BCD, QMD, and Drude contributions in the given channels.}
\label{tab:chi_selection_rule}
\vspace{0.2cm}
\setlength{\tabcolsep}{10pt} 
\begin{tabular}{ll|ccc}
\hline
\hline
Channel & Response & BCD & QMD & Drude \\
\hline
$\sigma_{xyy}$ & $j_x \propto E_y^2$ & $\checkmark$ & $0$ & $0$ \\
\hline
$\sigma_{yxx}$ & $j_y \propto E_x^2$ & $0$ & $\checkmark$ & $\checkmark$ \\
\hline
\hline
\end{tabular}
\end{table}

The momentum-space distribution of the QMD kernel is shown in Fig.~\ref{fig:QMD_kernel}.
It is an even function of $k_x$ and, unlike the BCD density in Fig.~\ref{fig:BC_QM}(b)-(c),
remains finite at $m_z = 0$. Then, we illustrate the chemical potential ($|\mu|$)
dependence of the QMD-induced NHC for various $m_z$ values in Fig.~\ref{fig:sigma_BCD}(c).
As expected, $\sigma_{yxx}^{\rm QMD}$ owns a finite magnitude at $m_z = 0$
and diverges as $\mu \rightarrow 0$. This divergence manifests the singular QM characteristics
of the band node presented in Sec.~\ref{Sec_IID}. This behavior confirms that the QMD is
the sole quantum-geometric channel in chiral-symmetric SDSMs, thereby realizing a symmetric,
isolated, and pure quantum-metric nonlinear Hall effect.
For a finite $m_z$, however, the QMD-induced Hall conductivity develops a distinct peak
that shifts to larger chemical potentials and diminishes in amplitude as $m_z$ increases.

To be brief, the Drude and QMD-induced nonlinear Hall responses are intrinsically present and
mixed in the $\sigma_{yxx}$ channel of the type-II SDSM as shown in Table~\ref{tab:chi_selection_rule}.
Conversely, the BCD-induced response appears solely in the $\sigma_{xyy}$ channel when $m_z \neq 0$,
allowing the BCD contribution to be measured independently. In conseqeunce, the primary remaining
challenge is to distinguish the mixed Drude and QMD contributions in $\sigma_{yxx}$. This motivates
our subsequent discussion on using scaling laws for their experimental separation.

\begin{figure}
\centering
\hspace{-1.1cm}
\includegraphics[width=0.79\linewidth]{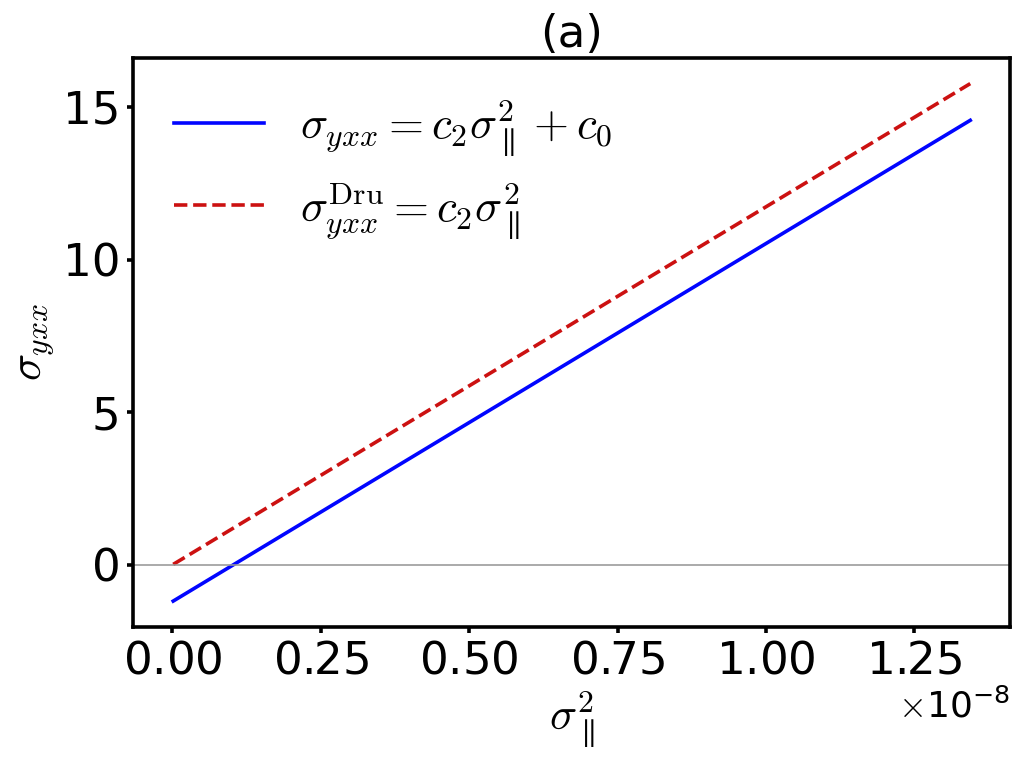}\\
\vspace{0.1cm}
\hspace{0.2cm}
\includegraphics[width=1.01\linewidth]{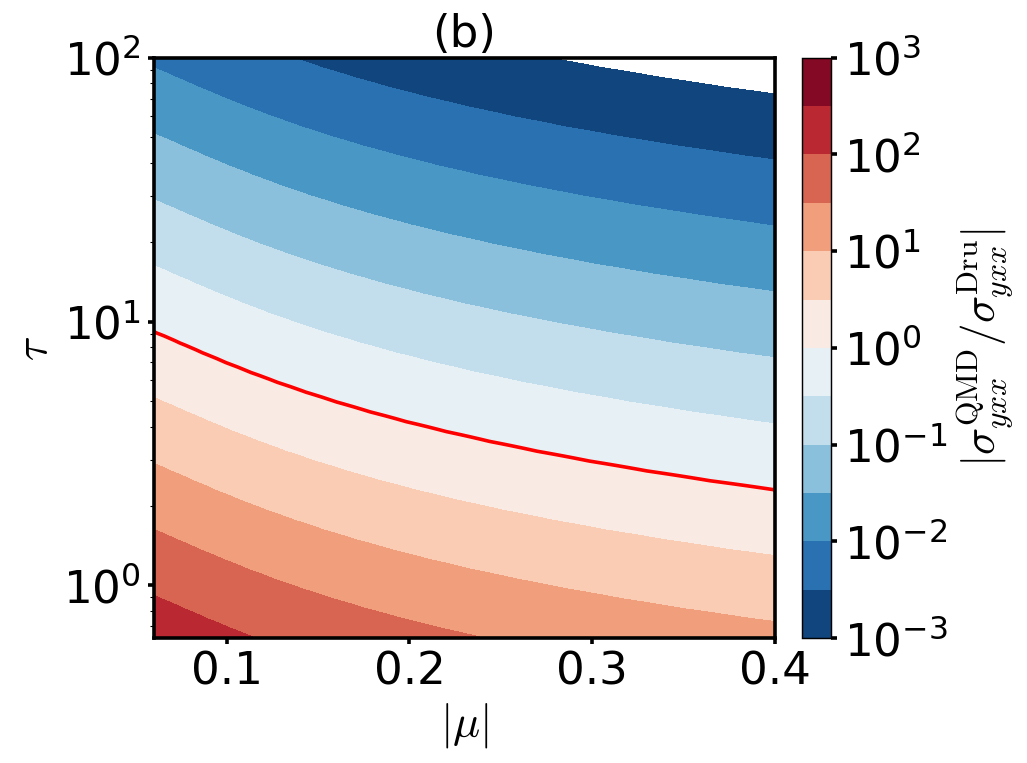}
\vspace{-0.39cm}
\caption{(Color online) (a) Scaling relation of the nonlinear Hall conductivity $\sigma_{yxx}$ versus $\sigma_\parallel^2$.
The blue solid line represents the total conductivity $\sigma_{yxx} = c_2 \sigma_\parallel^2 + c_0$,
while the red dashed line corresponds to the Drude contribution alone ($\sigma_{yxx}^{\rm Dru} = c_2 \sigma_\parallel^2$),
which intersects the coordinate origin. (b) Phase diagram of the ratio $|\sigma_{yxx}^{\rm QMD}/\sigma_{yxx}^{\rm Dru}|$ in the $(\tau, |\mu|)$ space, where the red curve indicates the critical boundary separating the Drude-dominated (above) and QMD-dominated (below) regimes. The parameters are chosen as $\gamma_1 = 0.75$, $\gamma_2 = 1.0$, $v = 0.65$, $m_z = 0.05$, and $\delta_0 = 0.01$.}
\label{fig:scaling}
\end{figure}


\subsection{Distinguishing Drude and QMD Contributions via Scaling Laws}

Since the Drude and QMD contributions coexist in the nonlinear Hall transport
and the relaxation time $\tau$ cannot be directly measured, it is necessary to distinguish
between them experimentally. Scaling laws provide a well-established approach for this task
and have been successfully applied in both linear and nonlinear transport regimes~\cite{Liu2025NSR,Gao2023SCI,Wang2023Nature,Gong2026SB,Nagaosa2010RMP,Tian2009PRL,
Hou2015PRL,Galitski2009PRB,Kaplan2024PRL,Fontana-Neto2025PRL}.

To derive the scaling law for the nonlinear Hall conductivity, we start from the
longitudinal Drude conductivity~\cite{Galitski2009PRB},
\begin{eqnarray}
\sigma_\parallel\equiv\sigma_{xx}=\frac{e^2\tau}{\hbar^2}S_{xx},
\end{eqnarray}
where $S_{xx} \equiv \sum_n \int_{\mathrm{BZ}} (\partial_{k_x} E_n)^2 \delta(E_n-\mu)$.
The relaxation time $\tau$ can then be formally expressed as
$\tau=\frac{\hbar^2}{e^2 S_{xx}}\sigma_\parallel$.
Inserting this into the Drude and QMD contributions yields the total conductivity of the \(yxx\) channel~\cite{Liu2025NSR,Gao2023SCI,Wang2023Nature,Gong2026SB,Kaplan2024PRL}:
\begin{eqnarray}
\sigma_{yxx}=c_2\sigma_\parallel^2+c_0,\label{eq:scaling}
\end{eqnarray}
where $c_0=\sigma_{yxx}^{\rm QMD}$ is the QMD term, and $c_2\equiv-\frac{\hbar}{e S_{xx}^2}\sum_n\int_{\mathrm{BZ}}(\partial_{k_x}^2 \partial_{k_y} E_n)f_n$ with $f_n$ defined in Eq.~(\ref{Eq_f}). By expressing the unmeasurable $\tau$ in terms of the experimentally observable $\sigma_\parallel$, this relation provides a direct criterion to separate the Drude and QMD contributions.

Equation~(\ref{eq:scaling}) indicates that in the absence of a QMD contribution ($c_0 = 0$),
the nonlinear Hall conductivity curve would passes directly through the coordinate origin,
as shown in Fig.~\ref{fig:scaling}(a). A finite QMD contribution, however, serves as
the $y$-intercept ($c_0$) of this curve.

Clearly, the competition between the two mechanisms reaches a critical threshold at $\sigma_\parallel^{*2}=|c_0|/|c_2|$.
In strongly disordered samples where $\sigma_\parallel < \sigma_\parallel^*$~\cite{Gao2023SCI,Gong2026PRX},
the QMD dominates the nonlinear Hall response. Conversely, in clean samples with $\sigma_\parallel > \sigma_\parallel^*$~\cite{Gao2023SCI,Gong2026PRX}, the Drude transport prevails. Figure~\ref{fig:scaling}(b)
presents the phase diagram of the ratio $|\sigma_{yxx}^{\rm QMD}/\sigma_{yxx}^{\rm Dru}|$ in the $(\tau, |\mu|)$
parameter space, with the red curve marking the critical boundary. As a result,
Drude transport dominates in the region above this curve (larger $\tau$),
whereas QMD dominates below it (smaller $\tau$).

In short, the scaling law in Eq.~(\ref{eq:scaling}) and the phase diagram in Fig.~\ref{fig:scaling}
offer a straightforward experimental strategy to distinguish between the Drude and QMD mechanisms in
future nonlinear transport experiments on type-II SD materials.

\section{Discussion and conclusion}\label{SecV}

\begin{figure}
\centering
\hspace{-0.25cm}
\includegraphics[width=\linewidth]{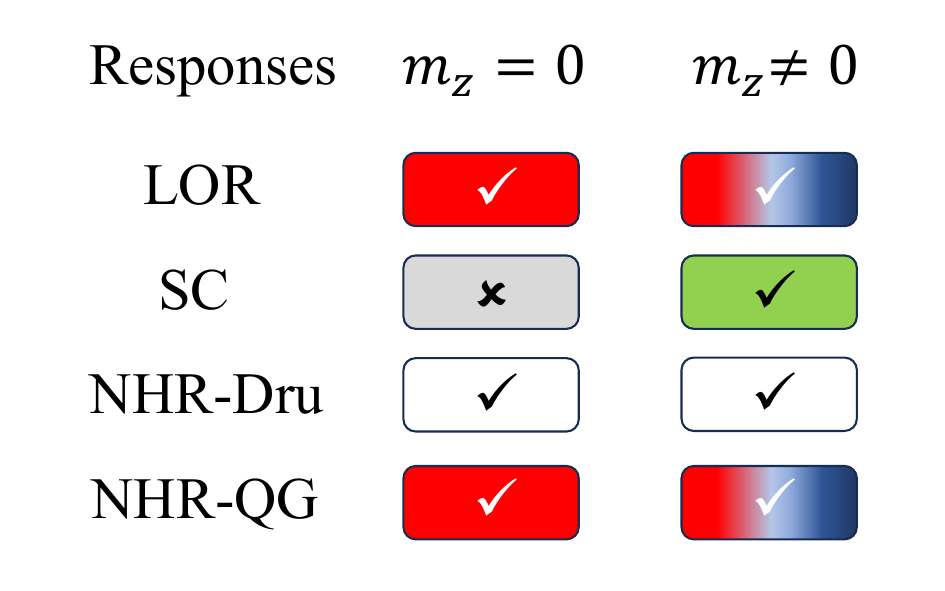}
\vspace{-0.35cm}
\caption{(Color online) Schematic summary of the tunable disentanglement of QG responses
for $m_z=0$ and $m_z \neq 0$. Rows correspond to the linear optical response (LOR),
shift current (SC), Drude nonlinear Hall response (NHR-Dru),and QG-induced nonlinear
Hall response (NHR-QG). The colors and symbols correspond respectively to pure QMD (red),
pure BCD (blue), mixed QMD and BCD (red-blue gradient), symmetry-allowed SC (green),
classical Drude transport (white), and no response (gray crosses).}
\label{fig:Schematic_Diagram}
\end{figure}

In this work, we have theoretically demonstrated a symmetry-protected, single-parameter pathway to cleanly isolate
the QM in type-II SDSMs as schematically summarized in Fig.~\ref{fig:Schematic_Diagram}.

For most platforms, QMD-induced nonlinear Hall responses are inevitably entangled with BCD contributions.
While this entanglement was recently circumvented using extrinsic antiferromagnetic symmetry engineering in $\mathrm{MnBi_2Te_4}$~\cite{Gao2023SCI,Wang2023Nature}, we show that the chiral limit of type-II SDSMs
intrinsically realizes this decoupling. Here, chiral symmetry strictly suppresses the BC, BCD, and the
shift current, isolating the QM as the sole intrinsic origin of the optical responses and the nonlinear
Hall effect as presented in Sec.~\ref{SecIII} and Sec.~\ref{SecIV}. These pure-QM responses are characterized
by a geometric singularity at the node, where the response diverges as $\mu \rightarrow 0$, and are
continuously tunable via the spin-orbit gap $m_z$.

The proposed mechanism is experimentally accessible in exfoliable type-II SD candidate materials,
including the $\mathrm{ZrSiS}$ family and 8-$\mathrm{Pmmn}$ borophene, where $m_z$ is continuously tunable by gate voltage or mechanical strain~\cite{Huang2015PRB,Elsayed2026PRR,Ghosh2026PRB,Bian2026PRB,Bezanilla2016PRB,Islam2018JPCM,Shao2024PRX,Sattari2025SR}. In particular, a finite $m_z$ in these materials breaks the chiral and $\mathcal{PT}$ symmetries,
leading to the coexistence of nonlinear Hall contributions from the Drude, BCD, and QMD terms.
By contrast, at the ideal chiral limit ($m_z=0$), the BCD term vanishes, leaving a mixture of
only the Drude and QMD contributions. This mixed response can be directly probed via the
second-harmonic Hall voltage $V_y^{2\omega}$~\cite{Gao2023SCI,Wang2023Nature},
and our derived scaling law, $\sigma_{yxx} = c_2\sigma_\parallel^2 + c_0$, allows
for the separation of these two remaining contributions. Specifically,
the $y$-intercept $c_0$ corresponds to the QMD, and the slope $c_2$ corresponds to the Drude term.
To cleanly extract the QMD contribution, strongly disordered samples are preferable, as the reduced $\tau$
suppresses the Drude response~~\cite{Gao2023SCI,Gong2026SB}.

In summary, the spin-orbit gap $m_z$ serves as a single parameter that drives the system between a pure-QM regime
and a mixed-geometry regime. In the mixed regime ($m_z\neq 0$), the emergence of BC activates the linear
optical conductivity and nonlinear shift current, which display single peaks governed by the band gap $2|m_z|$. Conversely,
at the chiral limit ($m_z=0$), all BC-contributed channels vanish entirely. This leaves the QM as the
sole surviving geometric origin, manifesting as a purely QM-governed longitudinal optical conductivity
and a nonlinear Hall response dictated by the interplay of Drude and QMD terms.
By virtue of their clean symmetry isolation, gate-tunable chiral point, and dual optical/transport
fingerprints, chiral type-II SDSMs serve as highly tunable platforms to isolate, detect,
and manipulate intrinsic quantum geometric effects. These results offer a promising blueprint for
future studies to identify and engineer similar isolated quantum metric responses in a broader class of semi-Dirac systems.

\section*{ACKNOWLEDGEMENTS}

J.W. is supported by Tianjin Natural Science Foundation Project (25JCYBJC01640) and
was partially supported by the National Natural
Science Foundation of China under Grant No. 11504360.


%
%



\begin{thebibliography}{99}

\bibitem{Liu2025NSR}
T. Liu, X.-B. Qiang, H.-Z. Lu, X. C. Xie,
{Quantum geometry in condensed matter},
\href{https://doi.org/10.1093/nsr/nwae334}{Natl. Sci. Rev. {\bf 12}, nwae334 (2025)}.

\bibitem{Gao2026RMP}
A. Gao, N. Nagaosa, N. Ni, and S.-Y. Xu,
{Quantum geometry phenomena in condensed matter systems},
\href{https://doi.org/10.1103/f78t-ky69}{Rev. Mod. Phys. {\bf 98}, 035005 (2026)}.

\bibitem{Das2023PRB}
K. Das, S. Lahiri, R. B. Atencia, D. Culcer, and A. Agarwal,
{Intrinsic nonlinear conductivities induced by the quantum metric},
\href{https://doi.org/10.1103/PhysRevB.108.L201405}{Phys. Rev. B {\bf 108}, L201405 (2023)}.

\bibitem{Orenstein2021ARCMP}
J. Orenstein, J. E. Moore, T. Morimoto, D. H. Torchinsky, J. W. Harter, and D. Hsieh,
{Topology and Symmetry of Quantum Materials via Nonlinear Optical Responses},
\href{https://doi.org/10.1146/annurev-conmatphys-031218-013712}{Annu. Rev. Condens. Matter Phys. {\bf 12}, 247 (2021)}.

\bibitem{Provost1980CMP}
J. P. Provost and G. Vallee,
{Riemannian structure on manifolds of quantum states},
\href{https://doi.org/10.1007/BF02193559}{Commun. Math. Phys. {\bf 76}, 289 (1980)}.

\bibitem{Berry1984PRSA}
M. V. Berry,
{Quantal phase factors accompanying adiabatic
changes},
\href{https://doi.org/10.1098/rspa.1984.0023}{Proc. R. Soc. A {\bf 392} (1802), 45 (1984)}.

\bibitem{Sipe2000PRB}
J. E. Sipe and A. I. Shkrebtii,
{Second-order optical response in semiconductors},
\href{https://doi.org/10.1103/PhysRevB.61.5337}{Phys. Rev. B {\bf 61}, 5337 (2000)}.

\bibitem{Sodemann2015PRL}
I. Sodemann and L. Fu,
{Quantum nonlinear Hall effect induced by Berry curvature dipole in time-reversal invariant materials},
\href{https://doi.org/10.1103/PhysRevLett.115.216806}{Phys. Rev. Lett. {\bf 115}, 216806 (2015)}.

\bibitem{Gao2023SCI}
A. Gao, Y.-F. Liu, J.-X. Qiu, B. Ghosh, T. V. Trevisan, Y. Onishi, C. Hu, T. Qian, H.-J. Tien, S.-W. Chen, M. Huang, D. B\'erub\'e, H. Li, C. Tzschaschel, T. Dinh, Z. Sun, S.-C. Ho, S.-W. Lien, B. Singh, K. Watanabe, T. Taniguchi, D. C. Bell, H. Lin, T.-R. Chang, C. R. Du, A. Bansil, L. Fu, N. Ni, P. P. Orth, Q. Ma, and S.-Y. Xu,
{Quantum metric nonlinear Hall effect in a topological antiferromagnetic heterostructure},
\href{https://doi.org/10.1126/science.adf1506}{Science {\bf 381}, 181 (2023)}.

\bibitem{Wang2023Nature}
N. Wang, D. Kaplan, Z. Zhang, T. Holder, N. Cao, A. Wang, X. Zhou, F. Zhou, Z. Jiang, C. Zhang, S. Ru, H. Cai, K. Watanabe, T. Taniguchi, B. Yan, and W. Gao,
{Quantum-metric-induced nonlinear transport in a topological antiferromagnet},
\href{https://doi.org/10.1038/s41586-023-06363-3}{Nature {\bf 621}, 487 (2023)}.

\bibitem{Xiao2010RMP}
D. Xiao, M.-C. Chang, and Q. Niu,
{Berry phase effects on electronic properties},
\href{https://doi.org/10.1103/RevModPhys.82.1959}{Rev. Mod. Phys. {\bf 82}, 1959 (2010)}.

\bibitem{Das2024PRL}
K. Das, K. Ghorai, D. Culcer, and A. Agarwal,
{Nonlinear Valley Hall Effect},
\href{https://doi.org/10.1103/PhysRevLett.132.096302}{Phys. Rev. Lett. {\bf 132}, 096302 (2024)}.


\bibitem{Kang2025NP}
M. Kang, S. Kim, Y. Qian, P. M. Neves, L. Ye, J. Jung, D. Puntel, F. Mazzola, S. Fang, C. Jozwiak, A. Bostwick, E. Rotenberg, J. Fuji, I. Vobornik, J.-H. Park, J. G. Checkelsky, B.-J. Yang, and R. Comin,
{Measurements of the quantum geometric tensor in solids},
\href{https://doi.org/10.1038/s41567-024-02678-8}{Nat. Phys. {\bf 21}, 110 (2025)}.

\bibitem{Kim2025SCI}
S. Kim, Y. Chung, Y. Qian, S. Park, C. Jozwiak, E. Rotenberg, A. Bostwick, K. S. Kim, and B.-J. Yang,
{Direct measurement of the quantum metric tensor in solids},
\href{https://doi.org/10.1126/science.ado6049}{Science {\bf 388}, 1050 (2025)}.

\bibitem{Sala2025SCI}
G. Sala, M. T. Mercaldo, K. Domi, S. Gariglio, M. Cuoco, C. Ortix, and A. D. Caviglia,
{The quantum metric of electrons with spin-momentum locking},
\href{https://doi.org/10.1126/science.adq3255}{Science {\bf 389}, 822 (2025)}.

\bibitem{Yu2025NC}
H. Yu, X. Li, Y.-Q. Bie, L. Yan, L. Zhou, P. Yu, and G. Yang,
{Quantum metric third-order nonlinear Hall effect in a non-centrosymmetric ferromagnet},
\href{https://doi.org/10.1038/s41467-025-63133-7}{Nat. Commun. {\bf 16}, 7698 (2025)}.

\bibitem{Park2026PRB}
T. Park, X. Huang, L. Savary, and L. Balents,
{Quantum geometry from the Moyal product: Quantum kinetic equation and nonlinear response},
\href{https://doi.org/10.1103/x9pk-lb4x}{Phys. Rev. B {\bf 113}, 045146 (2026)}.

\bibitem{Zhu2025NC}
H. Zhu, J. Li, X. Chen, Y. Yu, and Q. Liu,
{Magnetic geometry induced quantum geometry and nonlinear transports},
\href{https://doi.org/10.1038/s41467-025-60128-2}{Nat. Commun. {\bf 16}, 4882 (2025)}.

\bibitem{Pardo2009PRL}
V. Pardo and W. E. Pickett,
{Half-Metallic Semi-Dirac-Point Generated by Quantum Confinement in TiO$_2$/VO$_2$ Nanostructures},
\href{https://doi.org/10.1103/PhysRevLett.102.166803}{Phys. Rev. Lett. {\bf 102}, 166803 (2009)}.

\bibitem{Banerjee2009PRL}
S. Banerjee, R. R. P. Singh, V. Pardo, and W. E. Pickett,
{Tight-Binding Modeling and Low-Energy Behavior of the Semi-Dirac Point},
\href{https://doi.org/10.1103/PhysRevLett.103.016402}{Phys. Rev. Lett. {\bf 103}, 016402 (2009)}.

\bibitem{Delplace2010PRB}
P. Delplace and G. Montambaux,
{Semi-Dirac point in the Hofstadter spectrum},
\href{https://doi.org/10.1103/PhysRevB.82.035438}{Phys. Rev. B {\bf 82}, 035438 (2010)}.

\bibitem{Banerjee2012PRB}
S. Banerjee and W. E. Pickett,
{Phenomenology of a semi-Dirac semi-Weyl semimetal},
\href{https://doi.org/10.1103/PhysRevB.86.075124}{Phys. Rev. B {\bf 86}, 075124 (2012)}.

\bibitem{Xiong2023PRB}
Q.-Y. Xiong, J.-Y. Ba, H.-J. Duan, M.-X. Deng, Y.-M. Wang, and R.-Q. Wang,
{Optical conductivity and polarization rotation of type-II semi-Dirac materials},
\href{https://doi.org/10.1103/PhysRevB.107.155150}{Phys. Rev. B {\bf 107}, 155150 (2023)}.

\bibitem{Sinha2020PRB}
P. Sinha, S. Murakami, and S. Basu,
{Quantum Hall studies of a Semi-Dirac Nanoribbon},
\href{https://doi.org/10.1103/PhysRevB.102.085416}{Phys. Rev. B {\bf 102}, 085416 (2020)}.

\bibitem{Yan2023PRB}
C.-X. Yan, C.-Y. Tan, H. Guo, and H.-R. Chang,
{Highly anisotropic optical conductivities in two-dimensional tilted semi-Dirac bands},
\href{https://doi.org/10.1103/PhysRevB.108.195427}{Phys. Rev. B {\bf 108}, 195427 (2023)}.

\bibitem{Elsayed2026PRR}
M. M. Elsayed, T. I. Lakoba, and V. N. Kotov,
{Interacting type-II semi-Dirac quasiparticles},
\href{https://doi.org/10.1103/6ktq-2nn6}{Phys. Rev. Research {\bf 8}, 033128 (2026)}.

\bibitem{Carbotte2019PRB}
J. P. Carbotte, K. R. Bryenton, and E. J. Nicol,
{Optical properties of a semi-Dirac material},
\href{https://doi.org/10.1103/PhysRevB.99.115406}{Phys. Rev. B {\bf 99}, 115406 (2019)}.

\bibitem{Oriekhov2022PRB}
D. O. Oriekhov and V. P. Gusynin,
{Optical conductivity of semi-Dirac and pseudospin-1 models: Zitterbewegung approach},
\href{https://doi.org/10.1103/PhysRevB.106.115143}{Phys. Rev. B {\bf 106}, 115143 (2022)}.

\bibitem{Huang2015PRB}
H. Huang, Z. Liu, H. Zhang, W. Duan, and D. Vanderbilt,
{Emergence of a Chern-insulating state from a semi-Dirac dispersion},
\href{https://doi.org/10.1103/PhysRevB.92.161115}{Phys. Rev. B {\bf 92}, 161115(R) (2015)}.

\bibitem{Ghosh2026PRB}
B. Ghosh, M. Bandyopadhyay, and S. Nandy,
{Quantum geometry-driven photogalvanic responses in semi-Dirac systems},
\href{https://doi.org/10.1103/jm68-sx23}{Phys. Rev. B {\bf 114}, 105113 (2026)}.

\bibitem{Bera2026arXiv}
K. Bera, A. Saha, D. Chowdhury,
{Quantum geometric signatures of Link-Unlink transitions and nonlinear Hall response in Hopf-link semimetals},
\href{https://doi.org/10.48550/arXiv.2608.25352}{arXiv:2608.25352 [cond-mat.mes-hall] (2026)}.

\bibitem{Galitski2009PRB}
V. Galitski and S. Sachdev,
{Paired electron pockets in the hole-doped cuprates},
\href{https://doi.org/10.1103/PhysRevB.79.134512}{Phys. Rev. B {\bf 79}, 134512 (2009)}.

\bibitem{Bian2026arXiv}
W.-H. Bian and J. Wang,
{Nontrivial topological properties from the cooperation of static perturbations and Floquet fields in type-II semi-Dirac semimetals},
\href{https://doi.org/10.48550/arXiv.2610.16659}{arXiv:2610.16659 [cond-mat.mes-hall] (2026)}.

\bibitem{Li2025arXiv}
Y. Li, Y. Liu, and C.-C. Liu,
Quantum metric-induced generalized magneto-optical effects in $\mathcal{PT}$-symmetric antiferromagnets,
\href{https://doi.org/10.48550/arXiv.2503.04312}{arXiv:2503.04312 [cond-mat.mtrl-sci] (2025)}.

\bibitem{Wiedmann2025PRB}
R. Wiedmann, K. Alpin, M. M. Hirschmann, and A. P. Schnyder,
{Quantum geometry and low-frequency optical conductivity of nodal planes},
\href{https://doi.org/10.1103/2f4l-tz7p}{Phys. Rev. B {\bf 112}, 125102 (2025)}.

\bibitem{Ezawa2024PRB}
M. Ezawa,
{Analytic approach to quantum metric and optical conductivity in Dirac models with parabolic mass in arbitrary dimensions},
\href{https://doi.org/10.1103/PhysRevB.110.195437}{Phys. Rev. B {\bf 110}, 195437 (2024)}.

\bibitem{Villegas2025PRB}
J. O. A. Biscocho and K. H. A. Villegas,
{Quantum geometric tensor in systems with fractional band dispersion},
\href{https://journals.aps.org/prb/abstract/10.1103/bymh-6r7r}
{Phys. Rev. B {\bf 111}, 235121 (2025).}

\bibitem{Uchoa-Elsayed2026arXiv}
B. Uchoa, M. M. Elsayed, V. N. Kotov, Y. Shao, and D. N. Basov,
{Colloquium: Semi-Dirac Fermions in Quantum Matter},
\href{https://doi.org/10.48550/arXiv.2607.20627}{arXiv:2607.20627 [cond-mat.str-el] (2026)}.

\bibitem{Bian2026PRB}
W.-H. Bian and J. Wang,
{Tunable phase transitions from semimetals to Chern insulators in two-dimensional quadratic-band-crossing materials},
\href{https://doi.org/10.1140/epjp/s13360-026-07583-y}{Eur. Phys. J. Plus {\bf 141}, 365 (2026)}.

\bibitem{Ahn-Nagaosa2020PRX}
J. Ahn, G.-Y. Guo, and N. Nagaosa,
{Low-Frequency Divergence and Quantum Geometry of the Bulk Photovoltaic Effect in Topological Semimetals},
\href{https://doi.org/10.1103/PhysRevX.10.041041}{Phys. Rev. X {\bf 10}, 041041 (2020)}.

\bibitem{Kwon-Yang2024PRB}
S. Kwon and B.-J. Yang,
{Quantum geometric bound and ideal condition for Euler band topology},
\href{https://doi.org/10.1103/PhysRevB.109.L161111}{Phys. Rev. B {\bf 109}, L161111 (2024)}.

\bibitem{Volovik2016arXiv}
G. E. Volovik,
{Topological Lifshitz transitions},
\href{https://doi.org/10.48550/arXiv.1606.08318}{arXiv:1606.08318 [cond-mat.other] (2016)}.

\bibitem{Volovik2018PU}
G. E. Volovik,
{Exotic Lifshitz transitions in topological materials},
\href{https://doi.org/10.3367/UFNe.2017.01.038218}{Phys. Usp. {\bf 61}, 89 (2018)}.

\bibitem{Matsuura2010PRB}
S. Matsuura and S. Ryu,
{Momentum space metric, nonlocal operator, and topological insulators},
\href{https://doi.org/10.1103/PhysRevB.82.245113}{Phys. Rev. B {\bf 82}, 245113 (2010)}.

\bibitem{Souza2000PRB}
I. Souza, T. Wilkens, and R. M. Martin,
{Polarization and localization in insulators: Generating function approach},
\href{https://doi.org/10.1103/PhysRevB.62.1666}{Phys. Rev. B {\bf 62}, 1666 (2000)}.

\bibitem{Resta2011EPJB}
R. Resta,
{The insulating state of matter: a geometrical theory},
\href{https://doi.org/10.1140/epjb/e2010-10874-4}{Eur. Phys. J. B {\bf 79}, 121 (2011)}.

\bibitem{Yu2025npjQM}
J. Yu, B. A. Bernevig, R. Queiroz, E. Rossi, P\"aiviT\"orm\"a, and B.-J. Yang,
{Quantum geometry in quantum materials},
\href{https://doi.org/10.1038/s41535-025-00801-3}{npj Quantum Mater. {\bf 10}, 101 (2025)}.

\bibitem{Roy2014PRB}
R. Roy,
{Band geometry of fractional topological insulators},
\href{https://doi.org/10.1103/PhysRevB.90.165139}{Phys. Rev. B {\bf 90}, 165139 (2014)}.

\bibitem{Jackson2015NC}
T. S. Jackson, G. M\"oller, and R. Roy,
{Geometric stability of topological lattice phases},
\href{https://doi.org/10.1038/ncomms9629}{Nat. Commun {\bf 6}, 8629 (2015)}.

\bibitem{Wang2021PRL}
J. Wang, J. Cano, A. J. Millis, Z. Liu, and B. Yang,
{Exact Landau Level Description of Geometry and Interaction in a Flatband},
\href{https://doi.org/10.1103/PhysRevLett.127.246403}{Phys. Rev. Lett. {\bf 127}, 246403 (2021)}.

\bibitem{Li2026PRL}
Y. Li and C.-C. Liu,
{Quantum-Metric-Based Optical Selection Rules},
\href{https://doi.org/10.1103/bdhy-hnd2}{Phys. Rev. Lett. {\bf 136}, 046901 (2026)}.


\bibitem{Biscocho2025PRB}
J. O. A. Biscocho and K. H. A. Villegas,
{Quantum geometric tensor in systems with fractional band dispersion},
\href{https://doi.org/10.1103/bymh-6r7r}{Phys. Rev. B {\bf 111}, 235121 (2025)}.

\bibitem{Li2026arXiv}
C.-A. Li, Y. Qin, B. Fu, J. Li,
{Hidden topology and strong quantum metric bounds in trivial systems},
\href{https://doi.org/10.48550/arXiv.2607.24208}{arXiv:2607.24208 [cond-mat.mes-hall] (2026)}.

\bibitem{Mera-Goldman2022SPP}
B. Mera, A. Zhang, and N. Goldman,
{Relating the topology of Dirac Hamiltonians to quantum geometry: When the quantum metric dictates Chern numbers and winding numbers},
\href{https://scipost.org/SciPostPhys.12.1.018}{SciPost Phys. {\bf 12}, 018 (2022)}.

\bibitem{Kolodrubetz2017PR}
M. Kolodrubetz, D. Sels, P. Mehta, and A. Polkovnikov,
{Geometry and non-adiabatic response in quantum and classical systems},
\href{https://doi.org/10.1016/j.physrep.2017.07.001}{Phys. Rep., {\bf 697}, 1 (2017)}.

\bibitem{Fuchs-Montambaux2010EPJB}
J. N. Fuchs, F. Piechon, M. O. Goerbig, and G. Montambaux,
{Topological Berry phase and semiclassical quantization of cyclotron orbits for two dimensional electrons in coupled band models},
\href{https://doi.org/10.1140/epjb/e2010-00259-2}{Eur. Phys. J. B {\bf 77}, 351 (2010)}.

\bibitem{Peotta2025book}
S. Peotta, K.-E. Huhtinen, and P. T\"orm\"a,
{Quantum geometry in superfluidity and superconductivity},
\href{https://doi.org/10.3254/ENFI250023}{Proceedings of the International School of Physics ``Enrico Fermi", {\bf 211}, 373 (2025)}.

\bibitem{Kubo1957JPSJ}
R. Kubo,
Statistical-Mechanical Theory of Irreversible Processes. I. General Theory and Simple Applications to Magnetic and Conduction Problems,
\href{https://journals.jps.jp/doi/10.1143/JPSJ.12.570}{J. Phys. Soc. Jpn. {\bf 12}, 570(1957)}.

\bibitem{Ahn2022NP}
J. Ahn, G.-Y. Guo, N. Nagaosa, and A. Vishwanath,
{Riemannian geometry of resonant optical responses},
\href{https://doi.org/10.1038/s41567-021-01465-z}{Nat. Phys. {\bf 18}, 290 (2022)}.

\bibitem{Aversa1995PRB}
C. Aversa and J. E. Sipe,
{Nonlinear optical susceptibilities of semiconductors: Results with a length-gauge analysis},
\href{https://doi.org/10.1103/PhysRevB.52.14636}{Phys. Rev. B {\bf 52}, 14636 (1995)}.

\bibitem{Young2012PRL}
S. M. Young and A. M. Rappe,
{First Principles Calculation of the Shift Current Photovoltaic Effect in Ferroelectrics},
\href{https://doi.org/10.1103/PhysRevLett.109.116601}{Phys. Rev. Lett. {\bf 109}, 116601 (2012)}.

\bibitem{Cook2017NC}
A. M. Cook, B. M. Fregoso, F. d. Juan, S. Coh, and J. E. Moore,
{Design principles for shift current photovoltaics},
\href{https://doi.org/10.1038/ncomms14176}{Nat. Commun. {\bf 8}, 14176 (2017)}.

\bibitem{Resta2024PRL}
R. Resta,
{Geometrical Theory of the Shift Current in Presence of Disorder and Interaction},
\href{https://doi.org/10.1103/PhysRevLett.133.206903}{Phys. Rev. Lett. {\bf 133}, 206903 (2024)}.

\bibitem{Dressel2002book}
M. Dressel and G. Gr\"uner,
{Electrodynamics of Solids: Optical Properties of Electrons in Matter},
\href{https://doi.org/10.1017/CBO9780511606168}{Cambridge University Press (2002)}.

\bibitem{Giuliani2005book}
G. Giuliani and G. Vignale,
{Quantum Theory of the Electron Liquid},
\href{https://doi.org/10.1017/CBO9780511619915}{Cambridge University Press (2005)}.

\bibitem{Neto2009RMP}
A. H. C. Neto, F. Guinea, N. M. R. Peres, K. S. Novoselov, and A. K. Geim,
{The electronic properties of graphene},
\href{https://doi.org/10.1103/RevModPhys.81.109}{Rev. Mod. Phys. {\bf 81}, 109 (2009)}

\bibitem{Mahan2000book}
G. D. Mahan,
{Many-Particle Physics},
\href{https://link.springer.com/book/10.1007/978-1-4757-5714-9}{(Plenum, New York, 2000)}.

\bibitem{Altland2010book}
A. Altland and B. D. Simons,
{Condensed Matter Field Theory},
\href{https://doi.org/10.1017/CBO9780511789984}{Cambridge University Press (2010)}.

\bibitem{Chiu2026PRB}
P. M. Chiu,
Optical signatures of flat bands and anisotropic quantum geometry in magic-angle twisted bilayer graphene,
\href{https://doi.org/10.1103/4cct-pfgs}{Phys. Rev. B {\bf 113}, 205416 (2026)}.

\bibitem{Ghosh2024SA}
B. Ghosh, Y. Onishi, S.-Y. Xu, H. Lin, L. Fu, A. Bansil,
Probing quantum geometry through optical conductivity and magnetic circular dichroism,
\href{https://doi.org/10.1126/sciadv.ado1761}{Sci. Adv. {\bf 10}, eado1761 (2024)}.

\bibitem{Oh2016SA}
C.-g. Oh, S.-W. Kim, K. W. Kim, B. Monserrat, J.-W. Rhim,
{Mass-invariant universal optical conductivity from quantum geometry},
\href{https://doi.org/10.1126/sciadv.ady2033}{Sci. Adv. {\bf 12}, eady2033 (2026)}.

\bibitem{Shah2026arXiv}
M. Shah,
Floquet Engineering of Topological Phases and Magneto-Optical Response in a Driven d-wave Altermagnet,
\href{https://doi.org/10.48550/arXiv.2608.11192}{arXiv:2608.11192 [cond-mat.mes-hall] (2026)}.

\bibitem{Ezawa2024PRB2}
M. Ezawa,
{Intrinsic nonlinear conductivity induced by quantum geometry in altermagnets and measurement of the in-plane N\'eel vector},
\href{https://doi.org/10.1103/PhysRevB.110.L241405}{Phys. Rev. B {\bf 110}, L241405 (2024)}.

\bibitem{Okamoto2026CP}
S. Okamoto, A. Moreo, N. Nagaosa, S. S. P. Parkin,
{Berry curvature and field induced intrinsic anomalous Hall effect in an antiferromagnet FeTe},
\href{https://doi.org/10.1038/s42005-026-02751-z}{Commun. Phys. (2026)}.

\bibitem{Qian2023npjCM}
Z. Qian, J. Zhou, H. Wang, and S. Liu,
{Shift current response in elemental two-dimensional ferroelectrics},
\href{https://doi.org/10.1038/s41524-023-01026-3}{npj Comput Mater {\bf 9}, 67 (2023)}.

\bibitem{Yoshida2025PRB}
H. Yoshida and S. Murakami,
{Diverging shift current responses in the gapless limit of two-dimensional systems},
\href{https://doi.org/10.1103/PhysRevB.111.155402}{Phys. Rev. B {\bf 111}, 155402 (2025)}.


\bibitem{Gerbi2025AIPA}
Z. D. Gerbi,
{Electronic and optical properties of molybdenum disulfide (MoS$_2$) mono layer using density functional theory (DFT) calculations},
\href{https://doi.org/10.1063/5.0256769}{AIP Advances {\bf 15}, 025313 (2025)}.

\bibitem{Gu2023N}
H. Gu, Z. Guo, L. Huang, M. Fang, and S. Liu,
{Investigations of Optical Functions and Optical Transitions of 2D Semiconductors by Spectroscopic Ellipsometry and DFT},
\href{https://doi.org/10.3390/nano13010196}{Nanomaterials, {\bf 13}, 196 (2023)}.

\bibitem{Nielsen2023arXiv}
D. O. Nielsen, C. G. Van de Walle, S. T. Pantelides, R. D. Schrimpf, D. M. Fleetwood, and M. V. Fischetti,
{A first-principles approach to closing the ``$10-100$ eV gap" for charge-carrier thermalization in semiconductors},
\href{https://doi.org/10.48550/arXiv.2308.03893}{arXiv:2308.03893 [cond-mat.mtrl-sci] (2023)}.

\bibitem{Yan2018arXiv}
Z. Yan,
{Precise determination of critical points of topological phase transitions via shift current in two-dimensional inversion asymmetric insulators},
\href{https://doi.org/10.48550/arXiv.1812.02191}{arXiv:1812.02191 [cond-mat.mes-hall] (2018)}.

\bibitem{Baltz1981PRB}
R. v. Baltz and W. Kraut,
{Theory of the bulk photovoltaic effect in pure crystals},
\href{https://doi.org/10.1103/PhysRevB.23.5590}{Phys. Rev. B {\bf 23}, 5590 (1981)}.

\bibitem{Tan2016npjCM}
L. Z. Tan, F. Zheng, S. M. Young, F. Wang, S. Liu, and A. M. Rappe,
{Shift current bulk photovoltaic effect in polar materials---hybrid and oxide perovskites and beyond},
\href{https://doi.org/10.1038/npjcompumats.2016.26}{npj Comput Mater {\bf 2}, 16026 (2016)}.

\bibitem{Rangel2017PRL}
T. Rangel, B. M. Fregoso, B. S. Mendoza, T. Morimoto, J. E. Moore, and J. B. Neaton,
{Large Bulk Photovoltaic Effect and Spontaneous Polarization of Single-Layer Monochalcogenides},
\href{https://doi.org/10.1103/PhysRevLett.119.067402}{Phys. Rev. Lett. {\bf 119}, 067402 (2017)}.

\bibitem{Wang2025CPL}
E. Wang, Y. Wang, and S. Meng,
{Quantum Metric, Topology and Second Harmonic Generation},
\href{https://dx.doi.org/10.1088/0256-307X/42/4/047302}{Chin. Phys. Lett., {\bf 42}(4), 047302 (2025)}.

\bibitem{Jiang2025RPP}
Y. Jiang, T. Holder, and B. Yan,
{Revealing quantum geometry in nonlinear quantum materials},
\href{https://iopscience.iop.org/article/10.1088/1361-6633/ade454}{Rep. Prog. Phys. {\bf 88}, 076502 (2025)}.

\bibitem{Shibata2026arXiv}
J. Shibata,
{Second-order dc conductivity in the velocity-gauge Keldysh formalism: gauge-invariant decomposition into nonlinear Drude, Berry-curvature-dipole, and quantum-metric responses},
\href{https://doi.org/10.48550/arXiv.2606.22359}{arXiv:2606.22359 [cond-mat.mes-hall] (2026)}.

\bibitem{Gong2026SB}
Z.-H. Gong, Z. Z. Du, H.-P. Sun, H.-Z. Lu, and X. C. Xie,
{Scaling analysis of quantum geometry in second-order nonlinear transport},
\href{https://doi.org/10.1016/j.scib.2026.07.062}{Sci. Bull. {\bf 71}, 4422 (2026)}.

\bibitem{Kaplan2023SP}
D. Kaplan, T. Holder, and B. Yan,
{Unifying semiclassics and quantum perturbation theory at nonlinear order},
\href{https://scipost.org/10.21468/SciPostPhys.14.4.082}{SciPost Phys. {\bf 14}, 082 (2023)}.

\bibitem{Dressel2002AJP}
M. Dressel, G. Gruener, and G. F. Bertsch,
{Electrodynamics of Solids: Optical Properties of Electrons in Matter},
\href{https://doi.org/10.1119/1.1516200}{Am. J. Phys. {\bf 70}, 1269 (2002)}.

\bibitem{Ulrich2026PRB}
Y. Ulrich, J. Mitscherling, Laura Classen, and A. P. Schnyder,
{Quantum geometric origin of the intrinsic nonlinear Hall effect},
\href{https://doi.org/10.1103/4z8z-4kch}{Phys. Rev. B {\bf 113}, L201107 (2026)}.

\bibitem{Ahmad2026PRB}
S. Ahmad and F. Xue,
{Length¨Cvelocity gauge equivalence of quantum geometric nonlinear conductivity in the adiabatic DC limit},
\href{https://doi.org/10.1103/r23j-t8kt}{Phys. Rev. B {\bf 114}, 154428 (2026)}.

\bibitem{Nagaosa2010RMP}
N. Nagaosa, J. Sinova, S. Onoda, A. H. MacDonald, and N. P. Ong,
{Anomalous Hall effect},
\href{https://doi.org/10.1103/RevModPhys.82.1539}{Rev. Mod. Phys. {\bf 82}, 1539 (2010)}.

\bibitem{Tian2009PRL}
Y. Tian, L. Ye, and X. Jin,
{Proper Scaling of the Anomalous Hall Effect},
\href{https://doi.org/10.1103/PhysRevLett.103.087206}{Phys. Rev. Lett. {\bf 103}, 087206 (2009)}.

\bibitem{Hou2015PRL}
D. Hou, G. Su, Y. Tian, X. Jin, S. A. Yang, and Q. Niu,
{Multivariable Scaling for the Anomalous Hall Effect},
\href{https://doi.org/10.1103/PhysRevLett.114.217203}{Phys. Rev. Lett. {\bf 114}, 217203 (2015)}.

\bibitem{Kaplan2024PRL}
D. Kaplan, T. Holder, and B. Yan,
{Unification of Nonlinear Anomalous Hall Effect and Nonreciprocal Magnetoresistance in Metals by the Quantum Geometry},
\href{https://doi.org/10.1103/PhysRevLett.132.026301}{Phys. Rev. Lett. {\bf 132}, 026301 (2024)}.

\bibitem{Fontana-Neto2025PRL}
P. Fontana, V. Velasco, C. Niu, P. D. Ye, P. V. Lopes, K. E. M. de Souza, M. V. O. Moutinho, C. Lewenkopf, and M. B. Silva Neto,
{Quantum Geometry and the Electric Magnetochiral Anisotropy in Noncentrosymmetric Polar Media},
\href{https://journals.aps.org/prl/abstract/10.1103/7nxc-j62y}{Phys. Rev. Lett. {\bf 135}, 106602 (2025)}.



\bibitem{Gong2026PRX}
Z.-H. Gong, Z.-H. Wei, H.-Z. Lu, and X. C. Xie,
{Identifying geometric third-order nonlinear transport in disordered materials},
\href{https://doi.org/10.1103/515g-qjq8}{Phys. Rev. X 16, 031012 (2026)}.

\bibitem{Bezanilla2016PRB}
A. Lopez-Bezanilla and P. B. Littlewood,
{Electronic properties of 8-Pmmn borophene},
\href{https://doi.org/10.1103/PhysRevB.93.241405}{Phys. Rev. B {\bf 93}, 241405(R) (2016)}.

\bibitem{Shao2024PRX}
Y. Shao,
{Semi-Dirac Fermions in a Topological Metal}
\href{https://doi.org/10.1103/PhysRevX.14.041057}{Phys. Rev. X {\bf 14}, 041057 (2024)}.

\bibitem{Sattari2025SR}
F. Sattari and S. Mirershadi,
{Spin-dependent conductance in gapped 8-$Pmmn$ borophene-superlattice},
\href{https://doi.org/10.1038/s41598-025-12028-0}{Sci. Rep {\bf 15}, 26441 (2025)}.

\bibitem{Islam2018JPCM}
S. F. Islam,
{Magnetotransport properties of 8-Pmmn borophene: effects of Hall field and strain},
\href{https://doi.org/10.1088/1361-648X/aac8b3}{J. Phys. : Condens. Matter {\bf 30}, 275301 (2018)}.




\end{thebibliography}
\end{document}